\documentclass[prb,english,twocolumn,amsmath,amssymb,superscriptaddress,longbibliography,aps]{revtex4-2}

\usepackage{amsmath}
\usepackage{graphicx}
\usepackage{amssymb}
\usepackage{dsfont}
\usepackage{color}
\usepackage{ulem}
\usepackage[bookmarks=true,colorlinks,linkcolor=blue,urlcolor=blue,citecolor=blue]{hyperref}

\usepackage{hyperref}
\usepackage{tikz}
\usetikzlibrary{calc}

\usepackage{caption}
\usepackage{subcaption}
\begin{document}

\title{Anyonic phase transitions in the 1D extended Hubbard model with fractional 
statistics}

\author{Martin Bonkhoff}
\affiliation{Physics Department and Research Center OPTIMAS, University of Kaiserslautern-Landau, 67663 Kaiserslautern, Germany}
\affiliation{Theoretische Physik, Univ.~Hamburg, Jungiusstr. 9, 20255 Hamburg, Germany}

\author{Kevin J\"{a}gering}
\affiliation{Physics Department and Research Center OPTIMAS, University of Kaiserslautern-Landau, 67663 Kaiserslautern, Germany} 

\author{Shijie Hu}
\affiliation{Physics Department and Research Center OPTIMAS, University of Kaiserslautern-Landau, 67663 Kaiserslautern, Germany}
\affiliation{Beijing Computational Science Research Center, Beijing 100193, China}

\author{Axel Pelster}
\affiliation{Physics Department and Research Center OPTIMAS, University of Kaiserslautern-Landau, 67663 Kaiserslautern, Germany}

\author{Sebastian Eggert}
\affiliation{Physics Department and Research Center OPTIMAS, University of Kaiserslautern-Landau, 67663 Kaiserslautern, Germany}

\author{Imke Schneider}
\affiliation{Physics Department and Research Center OPTIMAS, University of Kaiserslautern-Landau, 67663 Kaiserslautern, Germany}

\date{\today}

\begin{abstract}
We study one-dimensional (1D) lattice anyons with extended Hubbard interactions at unit filling using bosonization and numerical simulations. 
The behavior can be continuously tuned from Bosonic to Fermionic behavior by adjusting the topological exchange angle $\theta$,
which leads to a competition of different instabilities.
We present the bosonization theory in presence of dynamic gauge fields, which predicts a phase diagrams of
four different gapped phases with distinct dominant correlations.
Advanced numerical simulations determine and analyze the exact phase transitions between Mott insulator, 
charge density wave, dimerized state, and Haldane insulator, all of which meet at a multi-critical line in the parameter space of  anyonic 
angle $\theta$, onsite interaction $U$, and nearest neighbor repulsion $V$.
Superfluid and pair-superfluid phases are stable in a region of small $V$.

\end{abstract}

\maketitle

{\it Introduction.}
For continuous one-dimensional (1D) systems
an exotic particle species interpolating between Bosons and Fermions 
was first proposed by Leinaas and Myrnheim  \cite{anyons,footnote}.
This concept of anyons
has received renewed interest with the implementation by so-called ``lattice anyons''  \cite{Eckardt2017,Goldman2014,Bukov2015, Keilmann2011, Greschner2015,Lange2017,Straeter2016,Tang2015,Lange2017PRA,Liu2018a,Bonkhoff2021,Bonkhoff2023,Goerg2019, Lienhard2020,Kwan2023} based on
experimental progress to create 
artificial gauge fields by dynamic manipulations and Floquet driving in ultra-cold gases
 \cite{Bloch2008,Aidelsburger2013,Miyake2013,Struck2013,prl21,Schweizer2019,Langen2015,Gross2017, Goerg2019,  Lienhard2020,Kwan2023} or by Raman assisted coupling \cite{Froelian2022,Chisholm2022}.  
The orginally proposed continuous 1D anyons \cite{anyons} correspond to the limit of  
low-filling of interacting lattice anyons \cite{Bonkhoff2021}.
However, for unit filling of one particle per site
strong lattice effects give rise to a completely different situation, 
where several broken symmetry phases are possible, which will be the topic of this Letter.
This is motivated by the observation that different instabilities have been established
for Fermions and Bosons in the 1D extended Hubbard model at unit filling.
In particular, the competition 
of on-site $U$ and nearest-neighbor $V$ interactions
leads to a possible dimerized (or bond ordered) phase for Fermions  \cite{Nakamura99,Sengupta2002, Sandvik2004, Barbiero2017}, while for Bosons and small anyonic exchange angle
topological string correlations in form of a Haldane
insulator (HI) 
are observed in an intermediate regime \cite{Torre2006, Berg2008,Ejima2014, Rossini2012}.
This invites the questions for anyons: What is the nature of an intermediate phase 
as the anyonic exchange phase $\theta$ is tuned continuously between Bosonic 
behavior for $\theta=0$ to Fermionic behavior for $\theta=\pi$?  
The goal of this work is to examine the
transitions between all different instablities and determine the dominant correlations
as a function of $\theta$, $U$ and $V$ at unit filling.

The 1D extended Hubbard model for anyons can be written in standard notation
\begin{equation}
{\hat H}\!=\!{\sum_{l}}\!\left[\!-\!J{\hat a}^{\dagger}_{l}{\hat a}_{l+1}^{\phantom{\dagger}}\!+\!{\rm H.c.}
\!+\!{\frac{U}{2}}{\hat n}_{l} ({\hat n}_{l}-1)\! +\! V  {\hat n}_{l} {\hat n}_{l+1} \right]\!. \label{anyons} 
\end{equation}
 The anyonic statistics enters via  
the exchange phase $\theta$ in the deformed commutation relations on different sites $l\!\neq\! l'$:   
${\hat a}_{l}{\hat a}_{l'}^{\dagger}\! -\! e^{i\theta\,{\rm sgn}(l-l')}{\hat a}_{l'}^{\dagger}{\hat a}_{l} \!=\!0 $.
Via a generalized Jordan-Wigner  transformation $\hat a_l\!=\!\hat b_l \exp(i\theta \sum_{j<l}\!\hat n_j)$~\cite{Keilmann2011} the anyonic model can be mapped to a bosonic one \mbox{$[{\hat b}_{l'}^{\dagger},{\hat b}_{l}]=0$} for $l\neq l'$ with density-dependent hopping
\begin{align}
\label{kinetictermboson}
{\hat {H}_{\rm kin}}=-J{\sum_{l}}\left[{\hat b}^{\dagger}_{l}{\hat b}_{l+1} e^{i\theta{\hat n}_{l}}+{\rm H.c.}\right] 
\end{align}
The on-site commutator $l\! =\!l'$ depends on the physical implementation, which is
typically spinless bosons restricted to a maximum of two particles per site  \cite{Greschner2015}.  The limit of $\theta\! =\! \pi$ corresponds to spinless "pseudo-Fermions" with the same on-site behavior but anti-commutation relations for $l\!\neq \!l'$
We  now derive the corresponding low-energy field theory which will allow to draw conclusions about the expected phase transitions. 

{\it Bosonization of gauge fields.}
For the constraint of maximally two particles per site at average unit filling, an exact representation can be used  in terms of spin-1 operators
$\hat{S}^z_l = -\hat{n}_l + 1$ which take on eigenvalues $-1, 0, 1$  \cite{Dalmonte2011}. The operations of the
transformed hopping operator are given in terms of spin operators
$\hat{S}^+_l|-1 \rangle = \sqrt{2}|0\rangle$ and $S^+_l|0\rangle = \sqrt{2}|1\rangle$ as
\begin{equation}
\label{kineticterm_spin-1}
{\hat {H}_{\rm kin}}=-\frac{1}{2}{\sum_{l}}\left[t(\hat{S}^z_l,\hat{S}^z_{l+1}){\hat S}^{-}_{l}{\hat S}_{l+1}^+ +{\rm H.c.}\right], 
\end{equation}
where the effective hopping is $t(-1,0) = J 2 e^{i \theta}$,  $t
(0,1) = J$, $t(-1,1)=J \sqrt{2} e^{i \theta}$ and $t(0,0)=J \sqrt{2}$.  A similar 
parametrization for interacting bosons has been used before  \cite{Schulz1986,Berg2008} 
with the simplification $t=J$.  Here we keep the full expression for $t$ to avoid an 
artificial spin-flip symmetry and implement the $\theta$-dependence, which can be written in 
terms of spin-operators as
\begin{align}
t(\hat{S}^z_l,\hat{S}^z_{l+1} )=& 
J\left[1+d(\theta) \hat S_{l}^{z}\right]\left[1+d(0) (\hat S_{l+1}^{z}-1 ) \right] 
\end{align}
with $d(\theta)=1-\sqrt{2}e^{i\theta}$. 
Following Refs.~\cite{Berg2008,Schulz1986},   we introduce  two spin-$\frac{1}{2}$ operators  $\hat S_{l}^{-}=\hat S_{1,l}^{-}+\hat S_{2,l}^{-}$ and $\hat S_{l}^{z}=\hat S_{1,l}^{z}+\hat S_{2,l}^{z}$ for each site. This results in  a spin-$\frac{1}{2}$ two-leg ladder model
\begin{align}
 H_{\rm kin}=&-\frac{1}{2} \sum_{l}\left[ \tilde{t}\left(\hat{S}^z_{2,l},\hat{S}^z_{2,l+1}\right) \hat{S}^{-}_{1,l} \hat{S}^{+}_{1,l+1} \right. \\
&+\left. \tilde{t}\left(\hat{S}^z_{2,l},\hat{S}^z_{1,l+1}\right) \hat{S}^{-}_{1,l} \hat{S}^{+}_{2,l+1} + 1\leftrightarrow 2 +\rm{H.c.}\right].\nonumber\label{Hkin}
 \end{align}
 with a prefactor  $\tilde{t}(\hat{S}^z_{i,l},\hat{S}^z_{j,l+1} )\equiv t(\hat{S}^z_{i,l}-1/2,\hat{S}^z_{j,l+1}+1/2 )$ for $i,j=1,2$.
 Note, that the hopping in chain 1 depends on the spin operators $\hat{S}^z_{2,l}$ in chain 2
and vice versa.
In that sense we may speak of a lattice gauge function (see Appendix \ref{appendA})
 \begin{align}
    \tilde{t}(\hat{S}^z_{i,l},\hat{S}^z_{j,l+1} )   \!=\! \tilde{J}\!+\!J_z\hat{S}^{z}_{i,l} \!+\! \bar{J}_z\hat{S}^{z}_{j,l+1}\!+\!J_{z z} \hat{S}^{z}_{i,l} \hat{S}^{z}_{j,l+1}
\end{align} 
with $\tilde{J}\!=\!\frac{J}{4} (1\!+\!\sqrt{2})(1\!+\!\sqrt{2} e^{i \theta}) $ and higher spin interaction constants $J_z\!=\!\frac{J}{2} (1\!+\!\sqrt{2})(1\!-\!\sqrt{2} e^{i \theta})$, $\bar{J}_z\!=\!\frac{J}{2} (1\!-\!\sqrt{2})(1\!+\!\sqrt{2} e^{i \theta})$, and $J_{zz}\!=\!J(1\!-\!\sqrt{2})(1\!-\!\sqrt{2} e^{i \theta})$.

For low energies, the Hamiltonian of the coupled spin chains is expressed in terms of dual bosonic fields $\Theta_{n}$ and $\Phi_{n}$ for each chain $n=1,2$, which are defined for a continuum variable 
normalized such that 
$[\Phi_n(x),\Theta_{n^\prime}(x^\prime)]=\frac{i }{2} \delta_{n,n^\prime} {\rm sgn} (x-x^\prime)$  \cite{Berg2008}.  More details are given in the Appendix \ref{appendB}.
The resulting Hamiltonian is  
\begin{align}
\hat{H}\simeq  &\int dx \Big[\sum_{\nu=+,-}\frac{u_\nu}{2} 
\left(K_\nu\left(\partial_x\Theta_\nu\right)^{2}+K_\nu^{-1}\left(\partial_x\Phi_\nu\right)^{2}\right)\nonumber \\
&+\Delta \left(\partial_{x} \Theta_{+}
\partial_{x} \Phi_{+}+ \partial_{x} \Theta_{-}
\partial_{x} \Phi_{-}\right)
\label{Hamiltonianbosonized}\\
&+g_1\cos (\sqrt{4\pi}\Phi_{+}) 
+ g_2 \cos (\sqrt{4\pi}\Phi_{-})
\nonumber \\ 
&+g_3 \cos(\sqrt{4\pi}\Theta_{-}) 
+g_4 \cos(\sqrt{4\pi}\Phi_{-}) \cos(\sqrt{4\pi}\Phi_{-}) 
\Big],
\nonumber
\end{align}
which we will discuss below.
Here $\Phi_{\pm}=\Phi_{1} \pm \Phi_{2}$ and $\Theta_{\pm} = (\Theta_{1}\pm \Theta_{2})/2$ are the fields for the total (+) and relative ($-$) densities and phases of the two spin-1/2 chains.  Terms which only affect conserved quantum numbers have been omitted.
The parameters in the Hamiltonian can be determined in a weak coupling expansion to first 
order in $U$, $V$, $J_z$, $\tilde J_z$  and $J_{zz}$
\begin{align}
K_{+}& =2\left[1 +\frac{6 V+ U-4 \text{Re} J_{zz}/\pi}{\text{Re}( \pi\tilde{J}-J_{zz}/2\pi)} \right]^{-1/2} \label{Kp}\\
K_{-}& =2 \left[1 +\frac{2 V- U-4 \text{Re} J_{zz}/\pi}{\text{Re}(\pi\tilde{J}-J_{zz}/2\pi)} \right]^{-1/2} \\
u_{\nu}&=2 a \text{Re}(\tilde{J}-J_{zz}/2\pi^2)/K_\nu \\
\Delta&=-\frac{ a }{ \pi}\text{Im}(J_z+\bar{J}_z)=\frac{2 J a }{\pi}\sin \theta\label{bilinearcouplings}\\
\frac{g_{1}}{a}&=\frac{1}{2}(2V-U)+\frac{2 \text{Re}(J_{zz})}{\pi}\label{gonethetanull},
\\
\frac{g_{2}}{a}&=\frac{1}{2}(U-2V)+\frac{2 \text{Re}(J_{zz}) }{\pi}\label{gtwothetanull},
\\
\frac{g_{3}}{a}&=-\text{Re}(\tilde{J})\pi\label{gthreethetanull},
\\
\frac{g_4}{a}&=V-\frac{\text{Re}(J_{zz})}{\pi}\label{gfourthetanull},
\end{align} 
where $a$ is the lattice constant.  The dependence on the anyonic angle $\theta$ via $\tilde J, J_z, \tilde J_z, J_{zz}$
comes from an operator product expansion of the density-dependent hopping terms as discussed
in the Appendix \ref{appendB}.  It must be noted here, that the mapping from bosons to spin operators already 
implies a finite interaction $J_{zz}\!\neq\!0$ even for $\theta\!=\!U\!=\!V\!=\!0$.  
Therefore, the weak coupling limit is never exact and the actual coupling constants will
quantitatively differ from the formulas given above.  Nonetheless, the operator content and the qualitative behavior with increasing 
$\theta$, $V$ and $U$ is robust.

For $\theta\!=\!0$ and $\tilde t\!=\!J$ we recover the structure and phases for Bosons 
as discussed in Ref.~\cite{Berg2008}.  For $\theta>0$ we notice the appearance of 
a characteristic anyonic current-density interaction 
$\Delta$ 
on the second line in the Hamiltonian in Eq.~(\ref{Hamiltonianbosonized}).
This term is present for any filling and leads to different left- and right-moving 
velocities   \cite{Bonkhoff2021}.  While the corresponding 
time-dependent correlation functions now show chiral behavior, 
remarkably the static mode expansions of the fields remain unaffected   \cite{Bonkhoff2021} so 
scaling dimensions and the renormlization behavior are not changed by $\Delta$.

\begin{figure}[t]
  \includegraphics[width=0.48\textwidth]{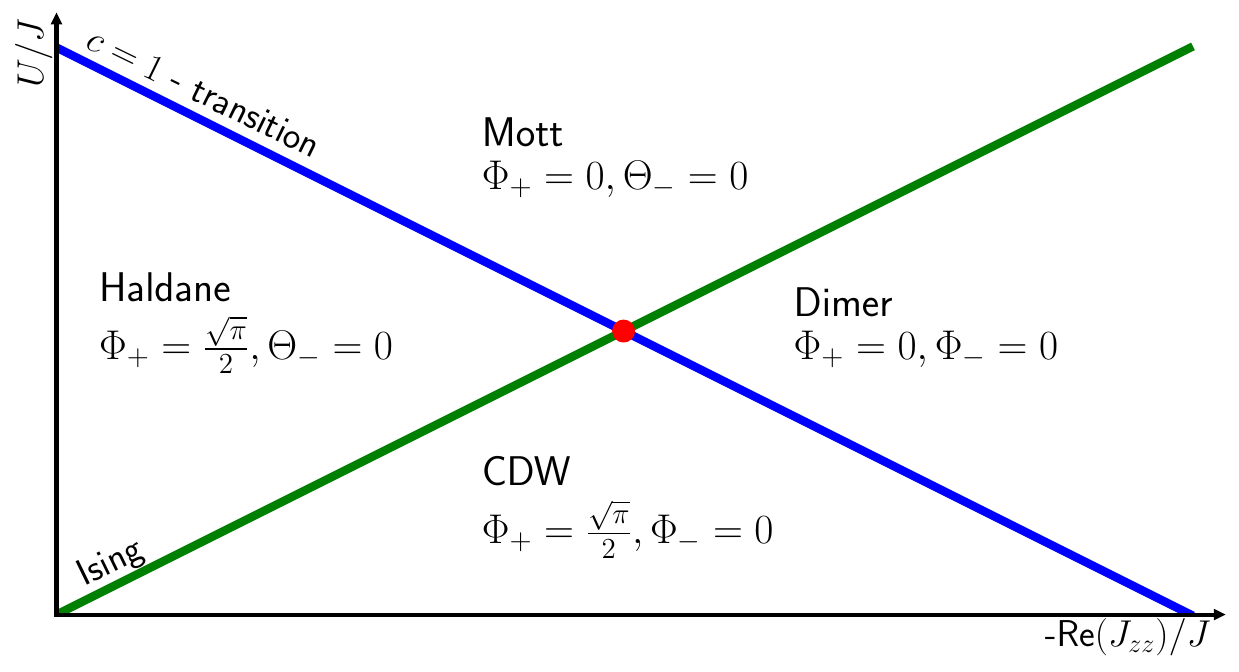}
  \vspace{-0.7cm}
  \caption{Schematic phase diagram resulting from Eqs.~(\ref{criticalU_G}) and (\ref{criticalU_I}) for fixed value of $V$ and $\tilde J$.} \vspace{-.5cm}
  \label{Paper6}
\end{figure}
{\it Phase transitions.}
Next, we turn to the $\cos$-interactions $g_1,g_2,g_3,g_4$ on the last two lines of the Hamiltonian (\ref{Hamiltonianbosonized}), all of which may be relevant and lead to 
gapped phases, depending on $K_\nu$   \cite{Berg2008}.  Notably, however,
the $\sin \sqrt{4\pi}\Phi_{+}$ term is absent, even though 
spatial inversion 
symmetry $\cal I$
is broken   \cite{Berg2008}.
This is due to the fact, that a modified inversion symmetry $\tilde{\mathcal I}$ is still obeyed
as discussed in Ref.~\cite{Lange2017}, which forbids this 
operator (see Appendix \ref{appendC} for a discussion of symmetries).

In the $(+)$~sector a pinning of the field $\Phi_+$ may occur due to the $g_1$-interaction,
which is relevant for all repulsive interactions $K_+<2$.   For $g_1<0$
the renormalization flow fixes the value of $\Phi_+=0$, while it becomes
$\Phi_+=\sqrt{\pi}/2$ for $g_1>0$, 
with a phase transition at $g_1=0$,
characterized by a free field of conformal charge $c=1$  \cite{Berg2008}. 
The relative quantum numbers in the 
$(-)$~channel will also be pinned with long-range correlations.  
Here, two interactions $g_2<0$ and $g_3<0$ compete with scaling dimensions
of $K_-$ and $1/K_-$, respectively.  The renormalization flow for $K_- <1$ will lead to 
a minimum at $\Phi_-=0$, while $K_->1$ is characterized by $\Theta_-=0$.
The phase transition for $K_-=1$ is predicted to be in the Ising universality class with $c=1/2$ as long as $g_2\approx g_3$ \cite{Berg2008, Lecheminant2002}.
For the model at hand a sign change of $g_2$ is only possible,
where it is not the leading instability.   
\begin{figure}[t]
	\includegraphics[width=0.48\textwidth]{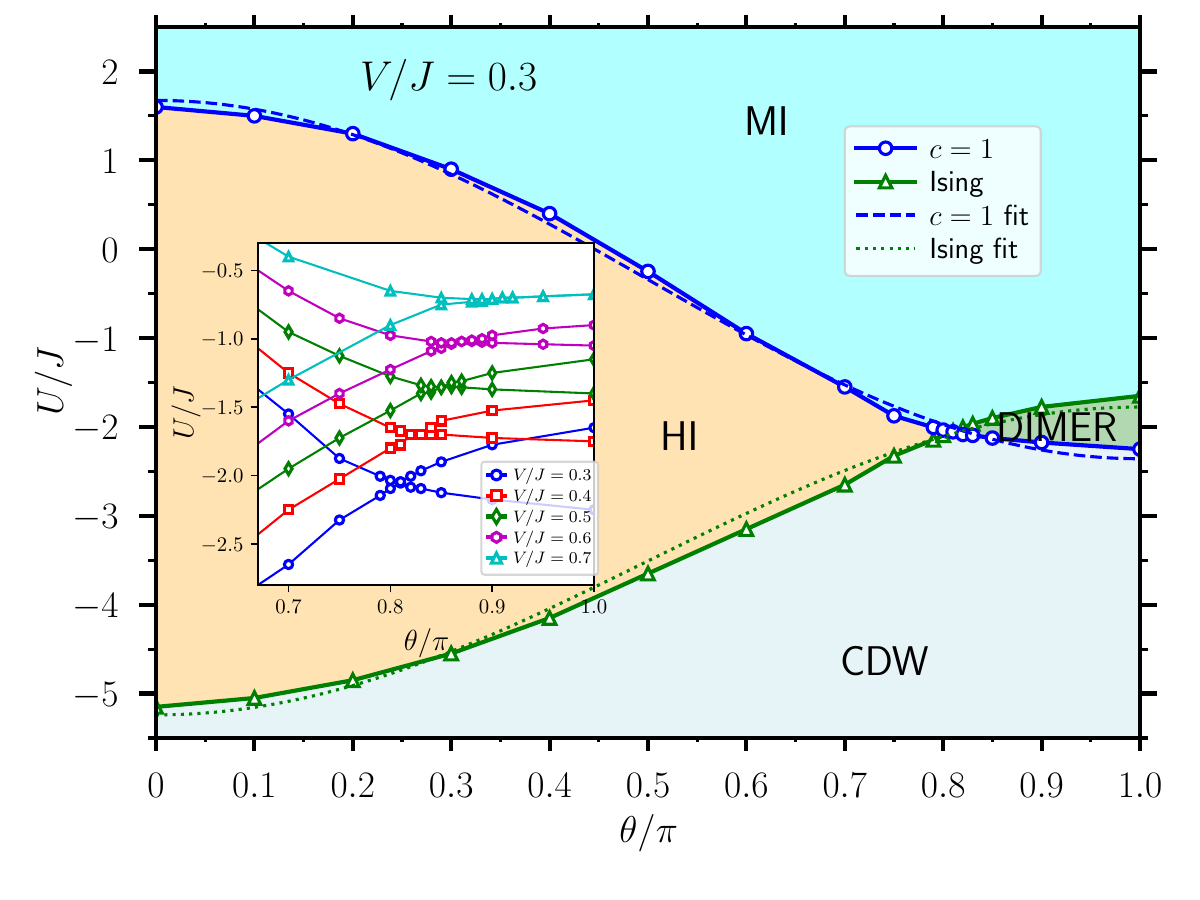}
	\vspace{-0.8cm}
	\caption{Phase diagram as a function of $\theta$ gained by iDMRG (M=400). Inset: Behavior for different $V$. } \vspace{-.5cm}
	\label{Paper11}
\end{figure}
The two phase transitions are plotted schematically in Fig.~\ref{Paper6} by 
using the conditions $g_1=0$ and $K_-=1$ to determine a critical $U^c$ as a function
of $-\text{Re}J_{zz}\propto 1-\sqrt{2}\cos{\theta}$ for some given value of $V$.
From Eqs.~(\ref{Kp})-(\ref{gfourthetanull}) we get in the weak-coupling limit
\begin{align}
U_{{c=1}}^c(\theta)&\sim 2V+\frac{4 \text{Re}[J_{zz}(\theta)]}{\pi} 
\label{criticalU_G}
\\
U_{\mathrm{Ising}}^c(\theta)&\sim 2V
-\frac{5 \text{Re}[J_{zz}(\theta)]}{2\pi} 
-3 \pi \text{Re}[\tilde{J}(\theta)] 
\label{criticalU_I}
\end{align}

For $\theta\!=\!0$ the corresponding long-range  ($|i-j|\!\to\! \infty$) order parameters 
have been discussed in Ref.~\cite{Berg2008} in terms of densities $\delta \hat{n}_l=\hat{n}_l-1$ and fields: 
For $U\!>\!U_{{c=1}}^c\!>\!U_{\mathrm{Ising}}^c$ the pinning  $\Phi_+\!=\!\Theta_-\!=\!0$ corresponds to
the Mott phase with parity order parameter 
\begin{equation}   
\mathcal{O}_{\rm MOTT}\!=\!\langle \mathrm{e}^{i \pi \sum_{i<k<j} \delta n_k } \rangle\! \sim\!\cos\sqrt{\pi}\Phi_+,
\label{OMOTT}\end{equation}
while $U_{{c=1}}^c\!>\!U_{\mathrm{Ising}}^c\!>\!U$  is a charge
density wave (CDW) with alternating charge order $\Phi_+\!=\!\sqrt{\pi}/2$ and fixed relative densities
$\Phi_-\!=0$ corresponding to the order parameter
\begin{equation}
  \mathcal{O}_{\rm CDW}\!=\! (-1)^{\vert i-j\vert}
\langle\delta\hat{n}_{i}\delta\hat{n}_ {j}\rangle\! \sim\!\sin\!\sqrt{\pi}\Phi_+\!\cos\!\sqrt{\pi}\Phi_-. \label{OCDW}
\end{equation}  In the intermediate
case, $U_{{c=1}}^c>U>U_{\mathrm{Ising}}^c$ a Haldane insulating (HI) phase has been 
identified  with alternating local charge $\Phi_+=\sqrt{\pi}/2$ but fluctuating 
relative densities $\Theta_-=0$ and topological string order parameter
\begin{equation}
\mathcal{O}_{\rm HI}=\langle \delta\hat{n}_i
\mathrm{e}^{i \pi \sum_{i\leq k<j} \delta\hat{n}_k} \delta\hat{n}_j \rangle\sim\sin \sqrt{\pi}\Phi_+.
\label{OHI}\end{equation}

Now for larger $\theta$,  anyons open  another possibility for the intermediate phase, as $U_{\mathrm{Ising}}^c$ increases and 
 $U_{{c=1}}^c$ decreases with $\theta$, such that $U_{{c=1}}^c\!<\!U\!<\!U_{\mathrm{Ising}}^c$, 
which is not possible for bosons.
In this case 
the order parameter is characterized by alternating 
energies on even and odd bonds
\begin{equation}
{\mathcal O}_{\mathrm{DIMER}}\!=\!\langle \hat{H}^{\mathrm{bond}}
_{\mathrm{even}}\! -\! \hat{H}^{\mathrm{bond}}_{\mathrm{odd}}\rangle
\!\sim\!\cos\!\sqrt{\pi}\!\Phi_+\!\cos\sqrt{\pi}
\Phi_-
\label{ODIMER}\end{equation} 
corresponding to uniform local charges $\Phi_+\!=\!0$ with relative quantum numbers
entangled on every second bond $\Phi_-\!=\!0$. For the Fermionic Hubbard model the dimer phase is
also known as a bond ordered wave  \cite{Sengupta2002}. However, it is not trivial that
such a phase is now also observed for $\theta=\pi$ 
corresponding to ''pseudo-Fermions''  
where the on-site commutator is bosonic with at most two particles per site. 
The Fermionic Hubbard model has a local Hilbert 
state of four states per site, while the model in Eq.~(\ref{anyons}) is restricted to three states, where the relative densities in the $(-)$ channel play the role of Fermionic spins.   

\begin{figure*}[t!]
	\begin{subfigure}{1\textwidth}	\includegraphics[width=0.245\textwidth]{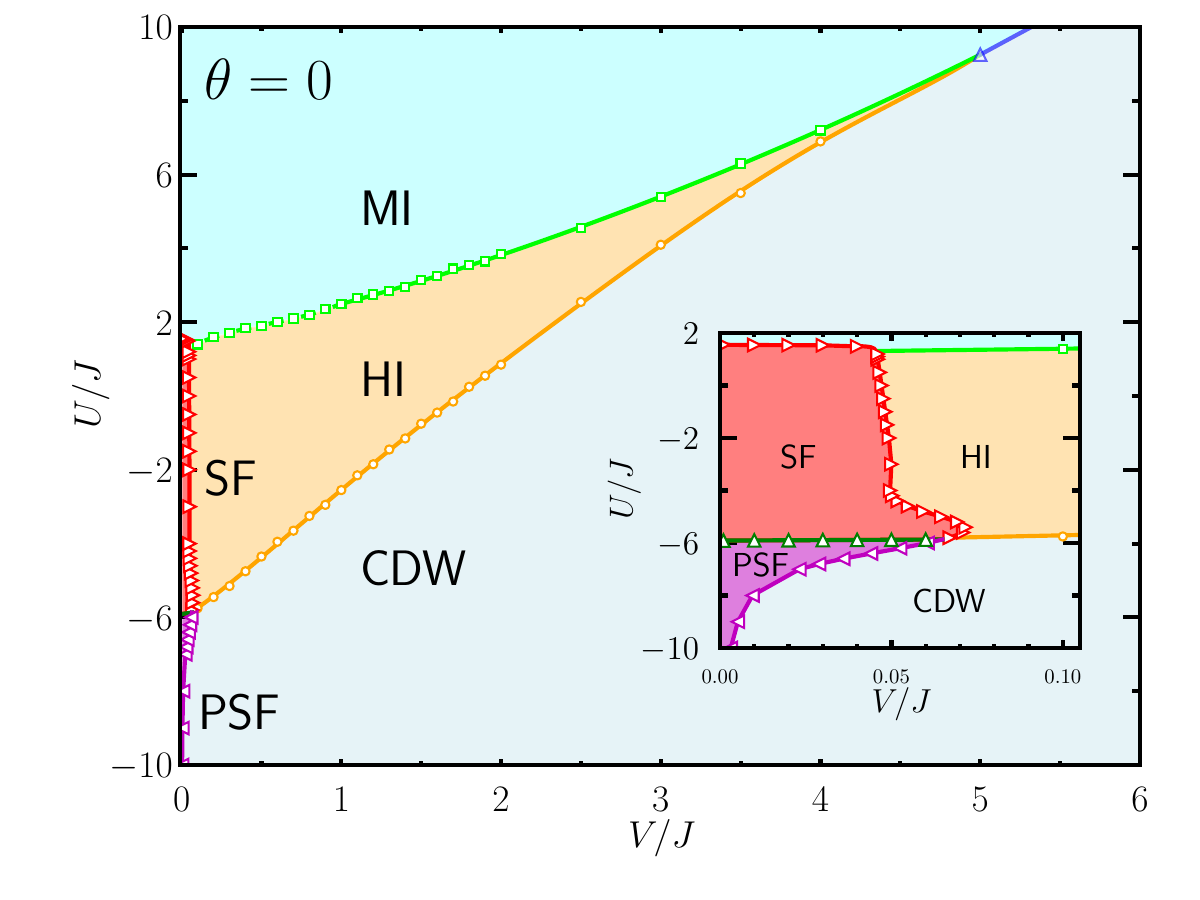} 
 \phantomsubcaption\label{PhasediagramTheta0}\includegraphics[width=0.245\textwidth]{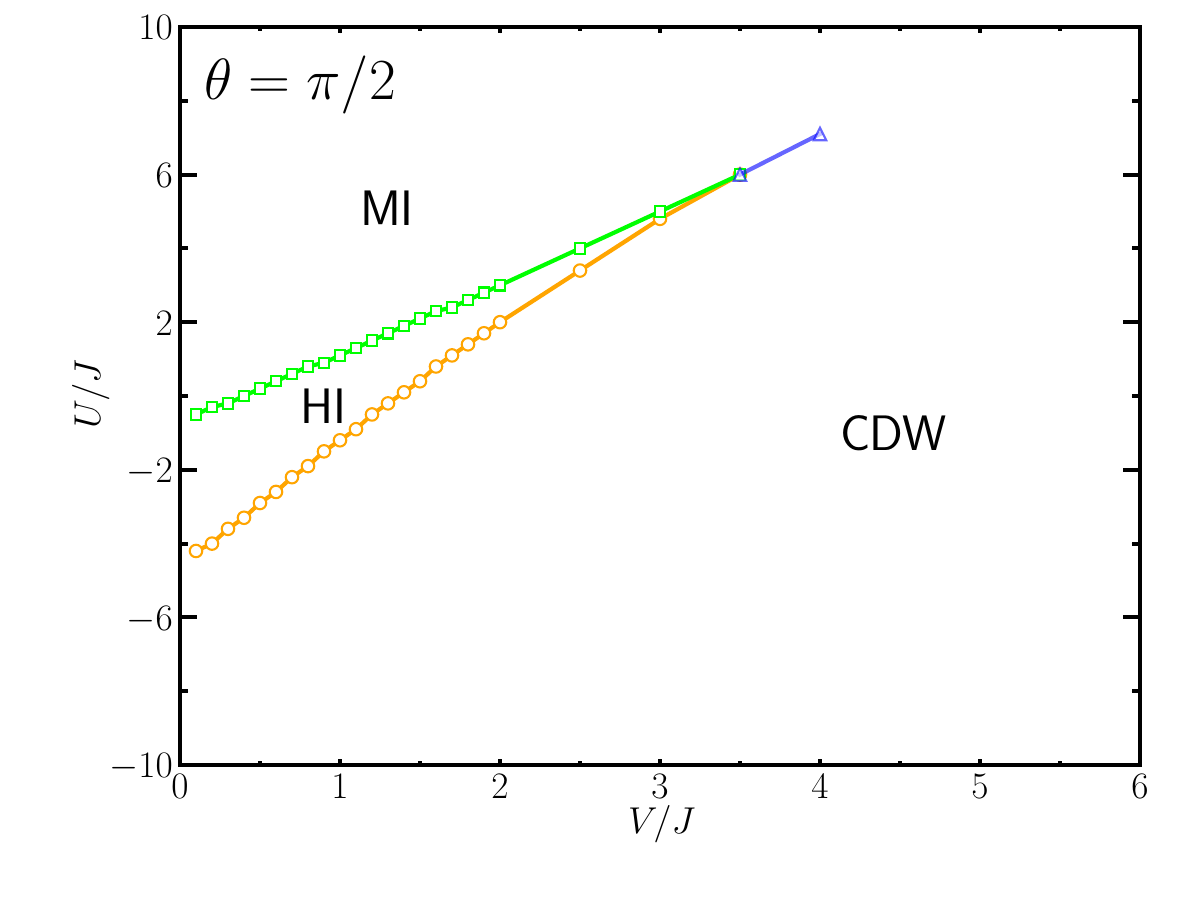}\phantomsubcaption\label{PhasediagramTheta0.5} 
		\includegraphics[width=0.245\textwidth]{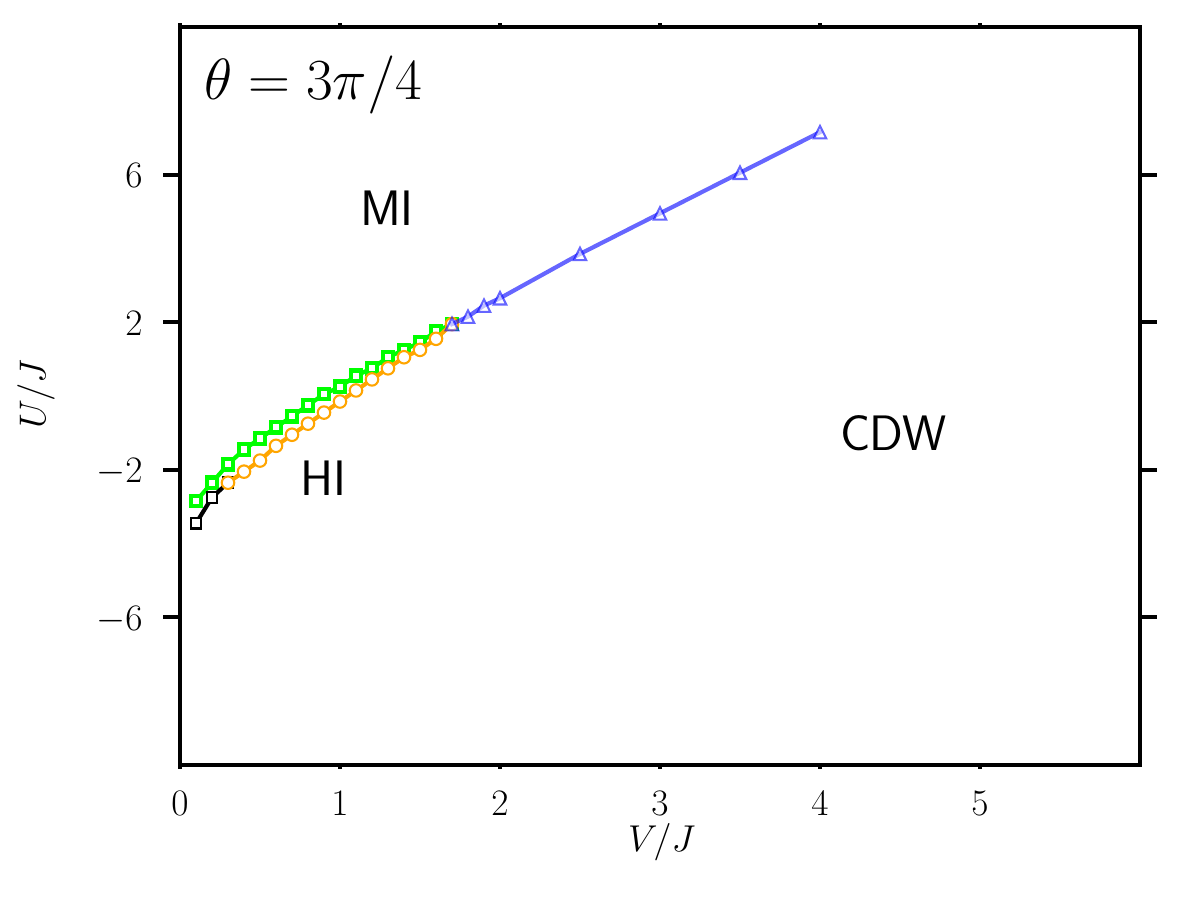}\phantomsubcaption\label{PhasediagramTheta0.75} 
		\includegraphics[width=0.245\textwidth]{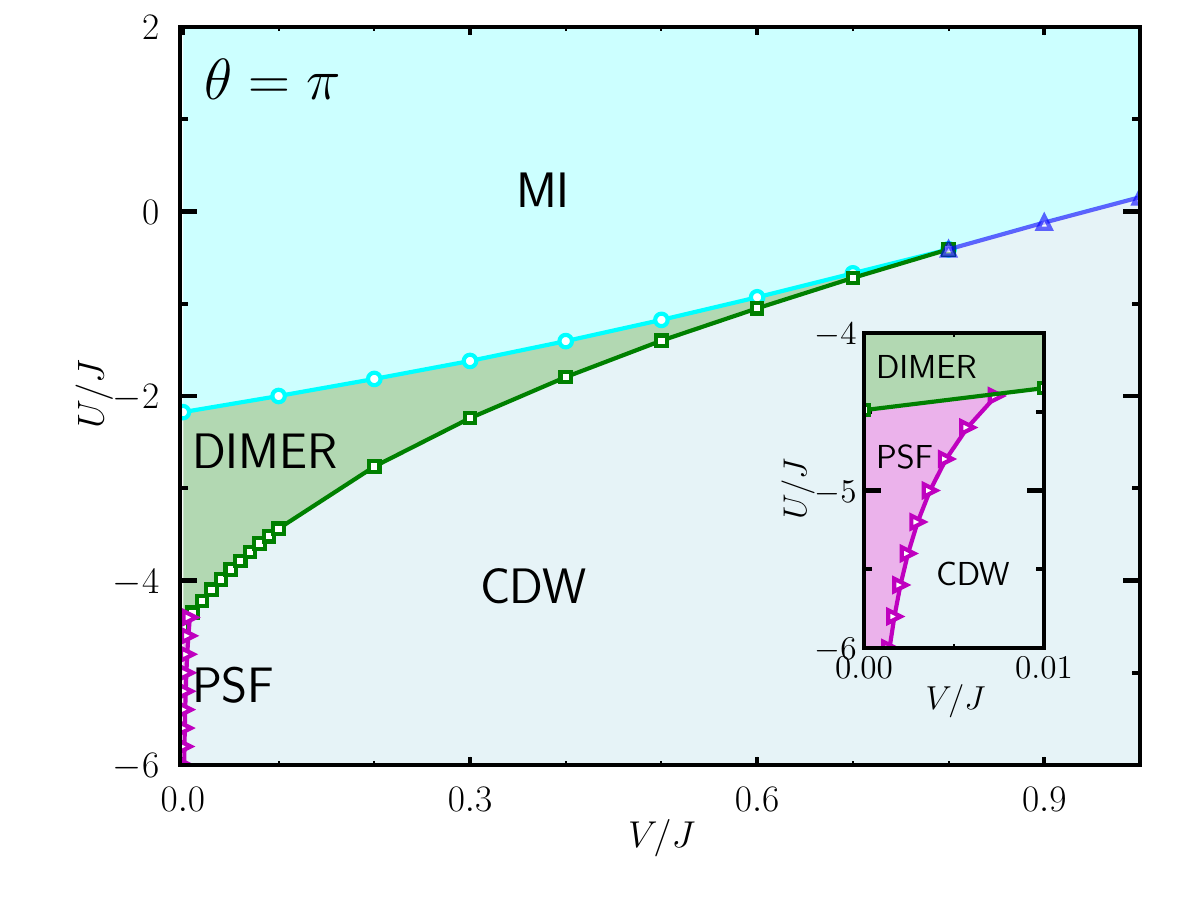}\phantomsubcaption\label{PhasediagramTheta1} 
	\vspace{-0.7cm}
	\end{subfigure}
	\caption{Phase diagram at $\theta=0,\theta=\pi/2,\theta=3\pi/4$ and $\theta=\pi$ determined by fidelity susceptibility, correlation length and entanglement peaks from iDMRG (M=100-800).} \vspace{-1.2cm}
	\label{Paper7}
\end{figure*}

{\it Numerical results.}
The density matrix renormalization group (DMRG) algorithm   \cite{White1992, White1993} and
the infinite DMRG (iDMRG)  \cite{Mcculloch2008} were used 
to calculate the fidelity susceptibility, the entanglement entropy, order parameters
and the correlation length, as well as level
crossing spectroscopy with finite-size scaling for accurate determination of the phase boundaries. The detailed analysis and data collapse is shown for selected points in the
Appendix \ref{appendD}, which yield the phase diagrams as shown in Fig.~\ref{Paper11} and Fig.~\ref{Paper7}
illustrating the fate of the HI phase for different values of $\theta$.
For $\theta=\pi/2$ the width of the HI phase has considerably decreased and for $\theta=3 \pi/4$ it is almost gone. 
For $\theta=\pi$ then the HI phase is completely replaced by a dimer  phase consistent 
with the prediction by bosonization. 
For the model at hand, the dimer phase extends to negative $U$ 
over a larger region than for the Fermionic Hubbard model  \cite{Sengupta2002}.

The bosonization results in Eqs.~(\ref{criticalU_G}) and (\ref{criticalU_I}) predict a $\cos\theta$ dependence
of the phase transition lines of the form $U_{{c=1}}^c(\theta)\!-\!U_{{c=1}}^c(\frac{\pi}{2}) \!= \!a  \cos\theta$ and
$U_{\mathrm{Ising}}^c(\theta) \!-\!  U_{\mathrm{Ising}}^c(\frac{\pi}{2}) \!=\! b  \cos\theta$, where the weak coupling expressions evaluate to $a\!\sim\!0.74$ and $b\!\sim\!-8.5$ with opposite sign.  For finite coupling values the prefactors change, but the cos-behavior appears to be 
robust as can be seen by the accurate
fits of the  phase transition lines in Fig.~\ref{Paper11}, that are given by 
$U_{c=1}^c\!=\!-0.34\!+\!2.02 \cos\theta$ and $U_{\mathrm{Ising}}^c\!=\!-3.51 \!-\! 1.73\cos\theta$.

The inset of Fig.~\ref{Paper11} shows the behavior near the points, where
all four phases meet for different $V$.
For each fixed $V$, the phase transition lines cross, which would signal the 
appearence of a four-critical point between the MI, HI, CDW, and dimer phases.
According to bosonization the multi-critical point would have central charge $c=3/2$ \cite{Ogino2021}.
All data is consistent with four phases meeting exactly at one multi-critical line in the $U-V-\theta$ parameter space.

More detailed phase diagrams in the $U-V$-parameter space for different $\theta$ are presented in Fig.~\ref{Paper7}.
An interesting aspect for small values of $V$ is the appearance of 
a superfluid (SF) for Bosons and pair-superfluid (PSF) phases for both Bosons and Pseudo-Fermions in the insets of Fig.~\ref{Paper7}.   An analytical description of those phases 
in terms of bosonization is beyond the scope of the work, however.  

\begin{figure}[t]
 \includegraphics[width=0.5\textwidth]{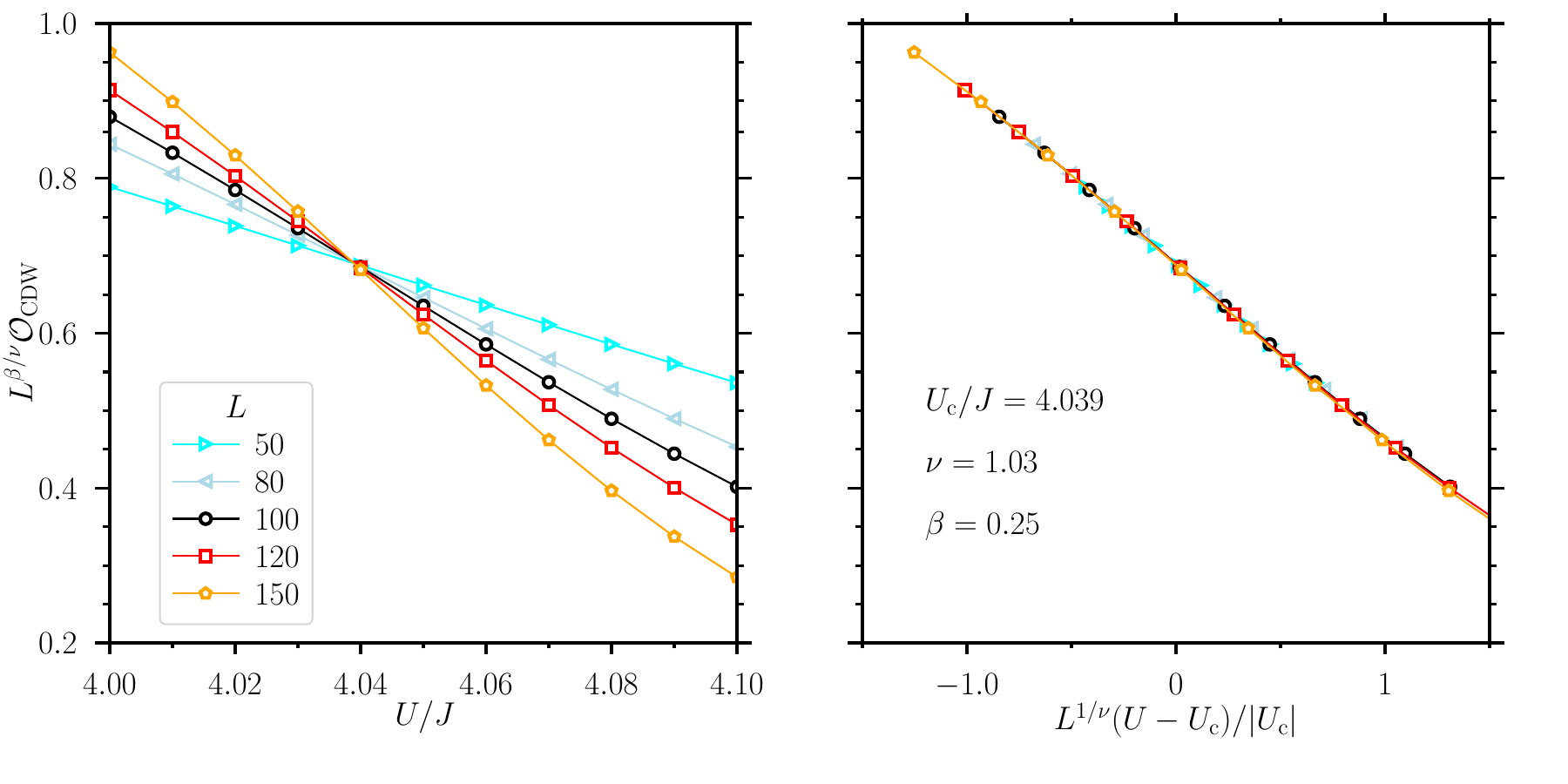}
  \includegraphics[width=0.5\textwidth]{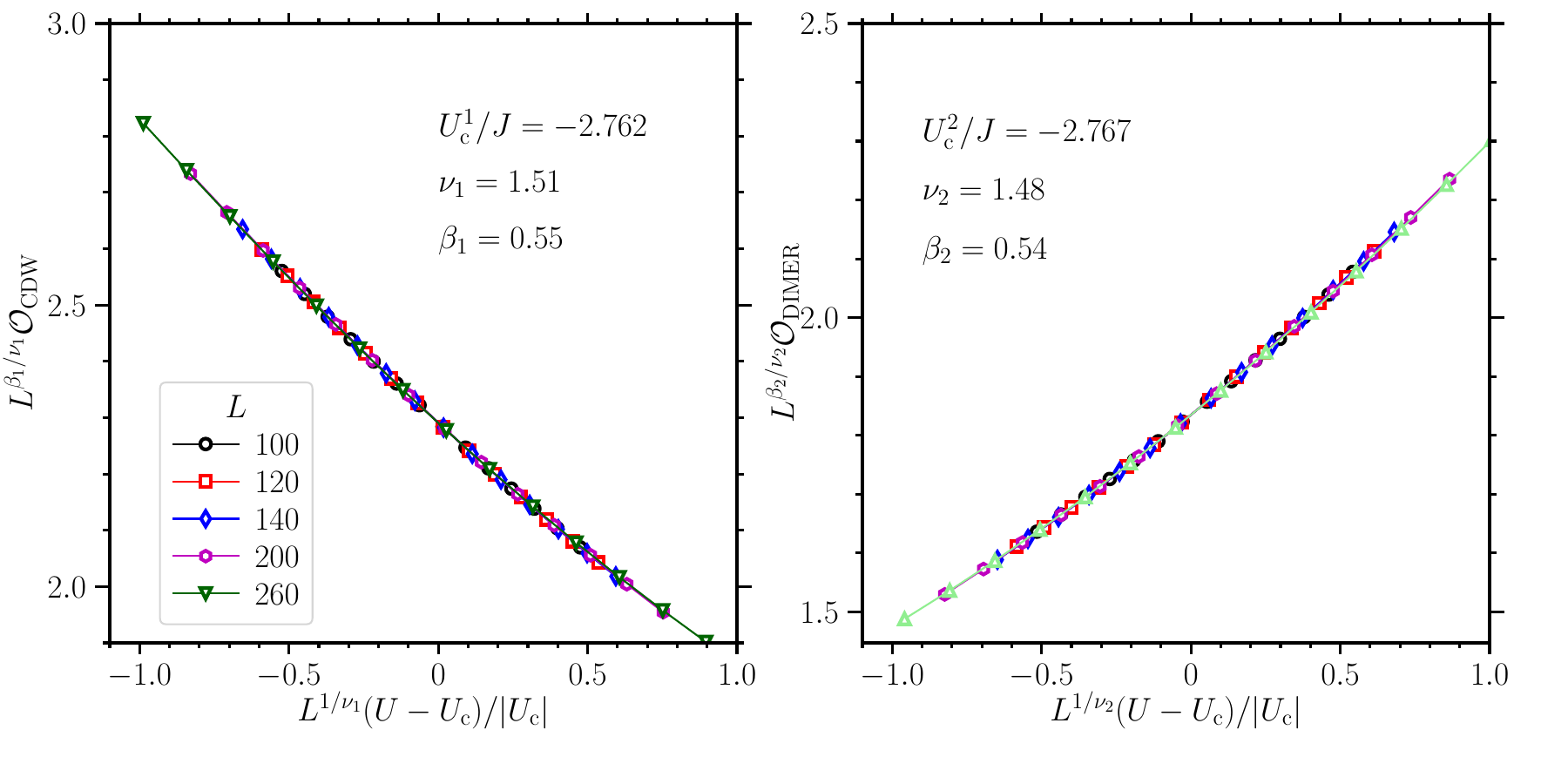}
  \vspace{-0.7cm}
  \caption{Data collapse of order parameters at the HI-CDW transition (upper: $V=3J,\theta=0$)
and the CDW-dimer transitions (lower: $V=0.2J, \theta=\pi$). Data from DMRG with $M=400$.}\vspace{-.5cm}
  \label{Paper5}
\end{figure}
{\it Classification of transitions.}
For the determination of universality classes the 
scaling of the order parameters is analyzed, 
which is numerically more involved.  This is illustrated in Fig.~\ref{Paper5}(upper)
for the transition between the  HI and CDW phases. The data collapse gives an accurate
determination of critical exponents $\nu=1.03$ and $\beta=.247$. 
This is in agreement  with the prediction of an Ising transition  
with $c=1/2$ \cite{Berg2008,Lange2017}, which we confirmed by entanglement entropy scaling  \cite{Nishimoto2011}
 also for the 
dimer to MI transition at large $\theta$ \ref{appendD}.

More interesting is the transition from CDW to dimer phase.
Both phases show a spontaneously broken translation symmetry, but different
broken inversion symmetries: bond centered and  site centered, respectively.
Recent literature has extensively discussed continuous transitions between
two phases with different broken ${\mathbb Z}_2$ symmetries as interesting examples 
 \cite{Macedo2022,
Mudry2019,Jiang2019, Huang2019, Roberts2019, Ogino2021, Luo2023, Zhang2023,
Senthil2004, Levin2004, Wang2017}
beyond the Landau-Ginzburg paradigm \cite{LandauLifshitz} in connection with 
deconfined quantum criticality  \cite{Senthil2004, Levin2004, Wang2017}.
For the anyon model we can confirm that the CDW-dimer transition also falls in 
this remarkable category.  It is continuous
with critical exponents close to $\nu=1.5$ and $\beta=0.5$ for a $c=1$ transition 
as shown in Fig.~\ref{Paper5}(lower).

For the extended Bose Hubbard model, the HI to Mott transition 
is known to be $c=1$  \cite{Ejima2014,Ejima2015,Rossini2012}. It is weaker for small
$V$  \cite{Lange2017} (see Appendix \ref{appendD}) and  merges to a first order
Mott-CDW transition for large $V$.
We observe
no significant change in the nature of the transition for finite $\theta$.  
A coupling in form of the operator $\sin \sqrt{4 \pi} \Phi_+$ is 
therefore always absent, which would 
otherwise lead to a crossover behavior instead of the entire $c=1$ line \cite{Berg2008}.

Finally, all transitions between SF/PSF and gapped phases appear to be of 
Berezinskii-Kosterlitz-Thouless type, independent of $\theta$. The corresponding transition 
lines were established  using level crossing spectroscopy starting with exact diagonalization of small systems to identify the relevant energy levels, followed by multi-targeting DMRG simulations up to $L=32$ extrapolated to the thermodynamic limit.

{\it Conclusions.}
By derivation of the effective field theory and  large scale numerical DMRG calculations 
we have  obtained the phase diagram as a function of $\theta$, $U$ and $V$ 
of the extended anyonic Hubbard model including negative onsite interactions.   
The phase diagram consists of four gapped and two superfluid phases.     
By tuning the exchange angle $\theta$ from 0 to $\pi$ the behavior changes continously 
from bosonic to fermionic.   In the process the Haldane insulator phase disappears
at an interesting multi-critical point where it is replaced by the dimer phase.
The transition between dimer and CDW phases provides an example of a
continous $c=1$ transition between two distinct broken ${\mathbb Z}_2$ symmteries beyond
the Ginzburg Landau paradigm.

\acknowledgements
We thank Nicholas Sedlmayr, Thore Posske, Sebastian Greschner, Pascal Jung, and Nathan Harshman for helpful discussions.  
We acknowledge support from
the Deutsche Forschungsgemeinschaft (DFG, German Research Foundation) 
Project No. 277625399-TRR 185 OSCAR (A4,A5,B6) and Forschungsgruppe FOR 2316 (project P10). 

\begin{widetext}

\appendix

\section{Lattice gauge function}
\label{appendA}
\label{sec:Peierls}

With the restriction of a maximum of two particles per site, the Bosonic creation operator can be expressed exactly in terms of spin-1 operators.
In the representation with triplet state of two spin-1/2 operators the final expression is given by
\begin{align}
\hat{b}_l^{\dagger} =&\frac{\hat S_{1,l}^{-}}{\sqrt{2}}  \left(a(0)+d(0) \hat S_{2,l}^{z}\right) +\frac{\hat S_{2,l}^{-}}{\sqrt{2}}  \left(a({0})+d({0})\hat S_{1,l}^{z}\right) 
\label{boscreation}
\end{align}
where $a(\theta)=(1+\sqrt{2}e^{i\theta})/2$ and $d(\theta)=(1-\sqrt{2}e^{i\theta})$. We will now derive how this representation is affected by the multiplication with a density dependent phase factor, i.e. we will derive the form of $\hat{b}^{\dagger}_{l} e^{i \theta \hat{n}_l }$.
The density-dependent phase decomposes as $e^{i\theta(\hat{n}_l-1)}=e^{-i\theta\hat{S}^z_{1,l}}e^{-i\theta\hat{S}^z_{2,l}}$ where the exponentials of spin-1/2 operators can be expressed in terms of trigonmetric functions
\begin{align}
e^{-i \theta{\hat S}_{n,l}^z}= \cos\left(\frac{\theta}{2}\right) \,{\rm id} - 2 i \sin \left(\frac{\theta}{2}\right) \hat{S}_{n,l}^z 
\label{exp_sz}
\end{align}
for $n=1,2$. We further notice that 
\begin{align}
    & \left(a({\tilde{\theta}}) +d({\tilde{\theta}})\hat S_{n,l}^{z}\right) \left(\cos\left(\frac{\theta}{2}\right)  - 2 i \sin \left(\frac{\theta}{2}\right)  \hat{S}_{n,l}^z \right) 
     = e^{-\frac{i \theta}{2} } \left(a({\tilde{\theta}+\theta}) + d({\tilde{\theta}+\theta}) \hat S_{n,l}^{z} \right)
\end{align}
for arbitrary angles $\tilde{\theta}$ and $\theta$.
Multiplication of the Boson operator with the density dependent phase therefore gives
\begin{align}
\hat{b}_l^{\dagger} e^{i \theta \hat{n}_l}=&\frac{\hat S_{1,l}^{-}}{\sqrt{2}}  \left(a(\theta)+d(\theta) \hat S_{2,l}^{z}\right) +\frac{\hat S_{2,l}^{-}}{\sqrt{2}}  \left(a(\theta)+d(\theta)\hat S_{1,l}^{z}\right) \enspace.
\label{spin_mapping_Peierls}
\end{align}
In conclusion,  on the level of spin-1/2 operators the action of the density dependent phase is solely reflected in the coefficients $a(\theta)$ and $d(\theta)$.  This expression also directly yields the lattice gauge function in Eq.~(6) of the main text.

\section{Bosonization}
\label{appendB}

\subsection{Mode expansion and Bosonization of spin operators }

\label{AppendixBosonization}
For the bosonization we follow the standard approach by first using the Jordan-Wigner mapping
\begin{align}
 \hat{S}_{q,l}^{+}=\hat{c}^{\dagger}_{q,l} \exp\left(i\pi \sum_{l^\prime<l} \hat{n}_{q,l^\prime}\right),\quad \hat{S}_{q,l}^z=\hat{n}_{q,l}-\frac{1}{2}
\end{align}
to a Fermionic representation where $\hat{c}^{\dagger}_{q,l}$ creates a Fermion on site $l$ in chain $q=1,2$ and $\hat{n}_{q,l}= \hat{c}^{\dagger}_{q,l} \hat{c}_{q,l}$. 
We then take the continuum limit and project the Fermionic fields onto states with momentum near the Fermi points $\pm k_{F}$.  This leads to right-
and left-moving components in the mode expansion:
\begin{equation}
\hat{c}_{q,l} \to \sqrt{a}\left(\hat{\Psi}^{q}_{R}(x)e^{ik_{F}x}+\hat{\Psi}^{q}_{L}(x)e^{-ik_{F}x}\right)
\end{equation}
where $x= a \,l$ and $a$ is the lattice constant. 
Next, we apply the standard Bosonization formula \cite{Giamarchi2003}
\begin{equation}
\hat{\Psi}^{q}_{R/L}(x)=\frac{1}{\sqrt{2\pi a}} e^{i \sqrt{\pi} (\Theta_{q}(x) \pm \Phi_{q}(x))} 
\label{BosonizationFermion}
\end{equation}
with the dual Bosonic fields $\Theta_{q}$ and $\Phi_{q}$ which fulfill $[\Phi_q(x),\Theta_{q^\prime}(x^\prime)]=\frac{i }{2} \delta_{q,q^\prime} {\rm sgn} (x-x^\prime)$. Here, the length scale $a$ also serves as the ultraviolet energy cutoff of the theory. The dual Bosonic fields can be decomposed into right- and left-moving fields $\Phi_{q}=\phi_{R,q}+\phi_{L,q}$ and $\Theta_q=\phi_{R,q}-\phi_{L,q}$.
We use the following mode expansions of these fields    
\begin{eqnarray}
\phi_{R,q}(x,t)&=&\phi_{R,q}^0 +\hat{Q}_{R,q}\frac{x-\upsilon t}{L}+\phi^+_{R,q}(x,t)+\phi_{R,q}^-(x,t)\\
\phi_{L,q}(x,t)&=&\phi_{L,q}^0 +\hat{Q}_{L,q}\frac{x+\upsilon t}{L}+\phi^+_{L,q}(x,t)+\phi_{L,q}^-(x,t)
\end{eqnarray} 
for finite system size $L$ and periodic boundary conditions  where
\begin{eqnarray}
\phi^-_{R,q}(x,t)&=&(\phi^+_{R,q}(x,t))^\dagger=\sum_n \frac{1}{\sqrt{4\pi n}} e^{i \frac{2\pi n}{L} (x-\upsilon t)} b_{R,n}^q\\\phi^-_{L,q}(x,t)&=&(\phi^+_{L,q}(x,t))^\dagger=\sum_n \frac{1}{\sqrt{4\pi n}} e^{-i \frac{2\pi n}{L} (x+\upsilon t)} b_{L,n}^q\enspace.
\end{eqnarray}
  The Bosonic modes $b_{R,n}^q$ and $b_{L,n}^q$ fulfill the canonical commutation relations $[b^{q}_{R/L,n},b^{q \,\dagger}_{R/L,n^\prime} ]=\delta_{n,n^\prime}$ if taken from the same branch, otherwise they commute.
The commutation rules  for the zero modes are  $[\phi_{R,q}^0,\phi_{L,q}^0]=\frac{i}{4}$ and $[ \phi^0_{R,q},\hat{Q}_{R,q}] =-\frac{i}{2}$ and $[ \phi^0_{L,q},\hat{Q}_{L,q} ] =\frac{i}{2}$ such that  the anti-commutation rules for the Fermion fields are ensured. For the Bosonic fields it follows
\begin{eqnarray}
[\phi_{R/L,q}^{-}(x,t),\phi_{R/L,q}^+(y,0)]&=&-\frac{1}{4\pi}\ln \left(1-e^{\pm i \frac{2\pi}{L} (x-y\,\mp\, \upsilon t )}e^{-\frac{2\pi}{L} a }\right) 
\end{eqnarray}
where we have again used the short distance cutoff $a$ to generate convergence. Using Baker-Cambell-Hausdorff $e^A e^B=e^{A+B}e^{\frac{1}{2}[A,B]}$ the vertex operators are normal-ordered: 
\begin{eqnarray}
\frac{1}{\sqrt{2\pi a}}\,e^{i \sqrt{4\pi}\phi_{R/L,q}(x)}=\frac{1}{\sqrt{L}}e^{i \sqrt{4\pi}\phi^{+}_{R/L,q}(x)}e^{i \sqrt{4\pi}\phi^{-}_{R/L,q}(x)} \enspace.
\end{eqnarray}
The spin operator $S^z_{q,l} = \hat{n}_{q,l}-1/2=:\hat{n}_{q,l}  :$ can now be bosonized using
\begin{align}
:\hat{n}_{q,l}  :&\simeq a\left[  \hat{\Psi}_{R}^{q\dagger}(x)\hat{\Psi}^q_{R}(x+a)+\,\hat{\Psi}_{L}^{q\dagger}(x)\hat{\Psi}^q_{L}(x+a)+(-1)^{\frac{x}{a}}\left(\hat{\Psi}_{R}^{q\dagger}(x)\hat{\Psi}^q_{L}(x)+\hat{\Psi}_{L}^{q \dagger}(x)\hat{\Psi}_{R}^q(x)\right)\right] \enspace.
\end{align}
For the operator product expansion of two Fermion operators of the same branch we find
\begin{align}
\hat{\Psi}_{ R/L}^{q \dagger}(x)\hat{\Psi}^q_{R/L}(x+a)=&
\frac{1}{L}:e^{ \mp i \sqrt{4\pi}\phi_{R/L,q}(x)}: :e^{ \pm i \sqrt{4\pi}\phi_{R/L,q}(x+a)}: \nonumber \\
\approx & \mp \frac{ i}{ 2\pi a} :e^{ \mp i \sqrt{4\pi}\left(\phi_{R/L,q}(x)-\phi_{R/L,q}(x+a)\right)}: \nonumber\\
= &  \frac{1}{2\pi a} \left[ \mp i + a \sqrt{4\pi}\, \frac{ \partial \phi_{R/L,q} (x)  }{\partial x} +  a^2 \sqrt{\pi}\, \frac{\partial^2 \phi_{R/L,q}(x)}{\partial x^2} \right. \nonumber \\
& \left. \pm i a^2 2\pi  \left(\frac{\partial \phi_{R/L,q}(x)}{\partial x}\right)^2 + \mathcal{O}(a^3)\right] \enspace 
\label{ope1}
\end{align}
whereas the rapidly oscillating product of two Fermion operators of different branches is given by
\begin{eqnarray}
(-1)^{\frac{x}{a}}\left(\hat{\Psi}_{R}^{q \dagger}(x)\hat{\Psi}_{L}^q(x)+\hat{\Psi}_{L}^{q\dagger}(x) \hat{\Psi}_{R}^q(x)\right)&=&\frac{(-1)^{\frac{x}{a}}}{2\pi a} e^{-2\pi [\phi^0_{R,q},\phi^0_{L,q}]} e^{-i \sqrt{4\pi}\Phi_q(x)}+\rm{H.c.}\\
&=&\frac{(-1)^{\frac{x}{a}}}{2\pi a} e^{-i\frac{\pi}{2}} e^{-i \sqrt{4\pi}\Phi_q(x)}+\rm{H.c.} \\
&=& - \frac{(-1)^{\frac{x}{a}}}{\pi a} \sin 2\sqrt{ \pi} \Phi_q(x). 
\label{osc1}
\end{eqnarray}
In leading order this results in 
\begin{align}
\hat{S}^z_{q,l}&\simeq \frac{a}{\sqrt{\pi}}\frac{\partial \Phi_q(x)}{\partial  x}-\frac{(-1)^{\frac{x}{a}}}{ \pi  }\mathrm{sin}2\sqrt{\pi}\Phi_q(x) \enspace .
\label{BosonizationspinSz}
\end{align}
For the spin raising operator we get
\begin{align}
\label{Bosonizationspin}
{\hat S}_{q,l}^{+}\simeq & \, e^{-i\sqrt{\pi}\Theta_q(x)}\left[b_1+ b_2 (-1)^{\frac{x}{a}}\sin\left(2\sqrt{\pi}\Phi_q(x)\right) \right]   
\end{align}
where $b_1$ and $b_2$ are non-universal constants. Note that the phases in the oscillating terms of Eqs.~(\ref{BosonizationspinSz}) and (\ref{Bosonizationspin}) depend on the treatment of zero modes where other conventions are possible  \cite{Giamarchi2003}.  

\subsection{Bosonization of hopping with dynamic gauge fields}
In this section we determine the bosonized expression for the correlated hopping including the density-dependent  phase. Our starting point is the spin-1/2 two-leg ladder model from Eq.~(5) in the main text 
\begin{align}
 H_{\rm kin}=&-\frac{1}{2} \sum_{l}\left[ \tilde{t}\left(\hat{S}^z_{2,l},\hat{S}^z_{2,l+1}\right) \hat{S}^{-}_{1,l} \hat{S}^{+}_{1,l+1} 
+  \tilde{t}\left(\hat{S}^z_{2,l},\hat{S}^z_{1,l+1}\right) \hat{S}^{-}_{1,l} \hat{S}^{+}_{2,l+1} + 1\leftrightarrow 2 +\rm{H.c.}\right]\nonumber\label{Hkin}
 \end{align}
 with the lattice gauge function
 \begin{align}
    \tilde{t}(\hat{S}^z_{i,l},\hat{S}^z_{j,l+1} )   &= \tilde{J}+J_z\hat{S}^{z}_{i,l} + \bar{J}_z\hat{S}^{z}_{j,l+1}+J_{z z} \hat{S}^{z}_{i,l} \hat{S}^{z}_{j,l+1}
\end{align}
for $i,j=1,2$
which we have derived in the main text with $\tilde{J}\!=\!\frac{J}{4} (1\!+\!\sqrt{2})(1\!+\!\sqrt{2} e^{i \theta}) $ and higher spin interaction constants $J_z\!=\!\frac{J}{2} (1\!+\!\sqrt{2})(1\!-\!\sqrt{2} e^{i \theta})$, $\bar{J}_z\!=\!\frac{J}{2} (1\!-\!\sqrt{2})(1\!+\!\sqrt{2} e^{i \theta})$, and $J_{zz}\!=\!J(1\!-\!\sqrt{2})(1\!-\!\sqrt{2} e^{i \theta})$. 
We proceed by first bosonizing the free hopping and the gauge function separately and then performing  operator-product expansions between the hopping and the lattice gauge terms whenever necessary.
As the lattice gauge function is complex-valued 
we thereby need to consider the operator product expansions between the hermitian conjugated pairs of the hopping and the lattice gauge explicitly.   After the Jordan-Wigner transformation the kinetic energy reads
 \begin{align}
 H_{\rm kin}=&-\frac{1}{2} \sum_{l}\left[ {t^\star}\left(\hat{n}_{2,l},\hat{n}_{2,l+1}\right) \hat{c}^{\dagger}_{1,l}\hat{c}_{1,l+1}^{\phantom{\dagger}}
+  t^\star\left(\hat{n}_{2,l},\hat{n}_{1,l+1}\right) \hat{c}^{\dagger}_{1,l} \hat{c}^{\phantom{\dagger}}_{2,l+1} + 1\leftrightarrow 2 +\rm{H.c.}\right]\nonumber\label{Hkinfermion}
\end{align}
where
\begin{align}
    t(\hat{n}_{i,l},\hat{n}_{j,l+1} )   &= \tilde{J}+J_z:\hat{n}_{i,l}: + \bar{J}_z:\hat{n}_{j,l+1}:+J_{z z} :\hat{n}_{i,l}: :\hat{n}_{j,l+1}:
\end{align} 
for $i,j=1,2$ and $^\star$ denotes complex conjugation while $:\ldots :$ stands for normal ordering by subtracting the ground state expectation value. 
The Fermionic bilinears for $k_F=\frac{\pi}{2}$ are given by
\begin{align}
\hat{c}^{\dagger}_{q,l} \hat{c}^{\phantom{\dagger}}_{q,l+1} \simeq  i\, a&\left[ \hat{\Psi}_{R}^{q\,\dagger}(x)\hat{\Psi}_{R}^q(x+a)-\,\hat{\Psi}_{L}^{q\,\dagger}(x)\hat{\Psi}_{L}^q(x+a) \right.\\ \nonumber
& \left. -(-1)^{\frac{x}{a}}\left(\hat{\Psi}_{R}^{q\,\dagger}(x)\hat{\Psi}_{L}^q(x+a)-\hat{\Psi}_{L}^{q\,\dagger}(x)\hat{\Psi}_{R}^q(x+a)\right)\right]
\end{align}
and are bosonized similarly to the local density in the previous subsection. In particular, we use Eq.~(\ref{ope1}) to determine the uniform contribution and proceed analogously to Eq.~(\ref{osc1}) for the oscillating contribution.  The result (before taking the Hermitian conjugate) reads
\begin{align}
	\label{bilinearappendix0}
	\hat{c}^{\dagger}_{q,l}\hat{c}^{\phantom{\dagger}}_{q,l+1}\simeq&\frac{1}{\pi}+i \frac{a}{\sqrt{\pi}}\frac{\partial \Theta_q(x)}{\partial x}-\frac{a^2}{2}\left[\left(\frac{\partial\Theta_q(x)}{\partial x}\right)^{2}+\left(\frac{\partial\Phi_q(x)}{\partial x}\right)^{2}\right]
		-\frac{(-1)^{\frac{x}{a}}}{\pi }\mathrm{cos}\left(2\sqrt{\pi}\Phi_q(x)\right).	
\end{align}
Note the different phase of the oscillating contribution compared to the local density. 
For the Bosonization of the gauge function we use $:\hat{n}_{q,l}:=\hat{S}^z_{q,l}$ in Eq.~(\ref{BosonizationspinSz})
and evaluate the product $:\hat{n}_{q,l}:\, :\hat{n}_{q,l+1}:$ by performing further operator product expansions. For the cross terms the following expressions are useful
\begin{align}
\frac{\partial\Phi_q(x)}{\partial x}\mathrm{sin}(2\sqrt{\pi}\Phi_q(x+a))\simeq &-\mathrm{sin}(2\sqrt{\pi}\Phi_q(x))\frac{\partial\Phi_q(x+a)}{\partial x}=\frac{1}{\sqrt{\pi}a}\mathrm{cos}(2\sqrt{\pi}\Phi_q(x))
\end{align}
and
\begin{align}
\sin\left(2\sqrt{\pi}\Phi_q(x)\right)\sin\left(2\sqrt{\pi}\Phi_q(x+a)\right)\simeq & \frac{1}{2}
- a^2 \pi \left(\frac{\partial \Phi_q(x)}{\partial  x}\right)^2 - \frac{1}{2}\cos\left(4\sqrt{\pi}\Phi_q\right)	\enspace.
\end{align}
Altogether, we find
\begin{align}
:\hat{n}_{q,l}:\, :\hat{n}_{q,l+1}:
&\simeq \frac{a^2}{\pi}\left(\frac{\partial \Phi_q(x)}{\partial  x}\right)^2+\frac{2 a (-1)^{\frac{x}{a}}}{\sqrt{\pi} \pi}\frac{\partial \Phi_q(x)}{\partial  x}\mathrm{sin}\left(2\sqrt{\pi}\Phi_q(x+a)\right)\\ \nonumber 
&-\frac{1}{\pi^2}\mathrm{sin}\left(2\sqrt{\pi}\Phi_q(x)\right)\mathrm{sin}\left(2\sqrt{\pi}\Phi_q(x+a)\right)\nonumber \\
&\simeq \frac{a^2}{\pi}\left(\frac{\partial \Phi_q(x)}{\partial  x}\right)^2+\frac{2  (-1)^{\frac{x}{a}}}{ \pi^2}\cos\left(2\sqrt{\pi}\Phi_q(x)\right) \\ \nonumber
&-\frac{1}{\pi^2}\left(\frac{1}{2}
- a^2 \pi \left(\frac{\partial \Phi_q(x)}{\partial  x}\right)^2 - \frac{1}{2}\cos\left(4\sqrt{\pi}\Phi_q\right)\right)
\end{align}
which can be summarized as
\begin{align}
:\hat{n}_{q,l}:\, :\hat{n}_{q,l+1}:&= -\frac{1}{2\pi^2} +\frac{2 a^2}{\pi}\left(\frac{\partial \Phi_q(x)}{\partial  x}\right)^2+\frac{2  (-1)^{\frac{x}{a}}}{ \pi^2}\cos\left(2\sqrt{\pi}\Phi_q(x)\right)
+ \frac{1}{2 \pi^2}\cos\left(4\sqrt{\pi}\Phi_q\right) \enspace.
\label{nnBosonized}
\end{align}

The different contributions to  the correlated intra-chain hopping 
\begin{align}\hat{\cal K}^{\text{intra}}_{l,l+1}&\equiv -\frac{1}{2}\left( {t^\star}\left(\hat{n}_{2,l},\hat{n}_{2,l+1}\right)\hat{c}^{\dagger}_{1,l}\hat{c}_{1,l+1}^{\phantom{\dagger}} + {t^\star}\left(\hat{n}_{1,l},\hat{n}_{1,l+1}\right)\hat{c}^{\dagger}_{2,l}\hat{c}_{2,l+1}^{\phantom{\dagger}}+ \rm{H.c.}\right) \\
&= \hat{\mathcal{K}}^a_{l,l+1}+\hat{\mathcal{K}}^b_{l,l+1}+\hat{\mathcal{K}}^c_{l,l+1}
\end{align}
can now be determined by direct multiplication. Here, we have defined
\begin{align}
   \hat{\mathcal{K}}^a_{l,l+1}\equiv& -\frac{1}{2} \sum_{q=1,2} \left(\tilde{J}^\star\hat{c}^{\dagger}_{q,l}\hat{c}_{q,l+1} + \rm{H.c.} \right) \\
   \hat{\mathcal{K}}^b_{l,l+1}\equiv & -\frac{1}{2} \sum_{q=1,2} \left[\hat{c}^{\dagger}_{q,l}\hat{c}_{q,l+1}\left(J_z^\star:\!\hat{n}_{\bar{q},l}\!:+\bar{J}_z ^\star:\!\hat{n}_{\bar{q},l+1}\!:\right) + \rm{H.c.} \right] \\
    \hat{\mathcal{K}}^c_{l,l+1}\equiv & -\frac{1}{2} \sum_{q=1,2} \left[ J_{z z}^\star\hat{c}^{\dagger}_{q,l}\hat{c}_{q,l+1} :\!\hat{n}_{\bar{q},l}\!:\,:\!\hat{n}_{\bar{q},l+1}\!:+ \rm{H.c.} \right]
\end{align}
where $\bar{q}=1$ if $q=2$ and $\bar{q}=2$ if $q=1$.
Since the evaluation of these terms only involves multiplication of fields with different chain indices we do not need to perform further operator product expansions.
We obtain
\begin{align}
\hat{\mathcal{K}}^a_{l,l+1} =&\sum_{q=1,2} \left\{\frac{  a^2  \text{Re}(\tilde{J})}{2}\left[\left(\partial_x\Theta_q\right)^{2}+\left(\partial_x\Phi_q \right)^{2}\right]  -\frac{ a\, \text{Im}(\tilde{J})}{ \sqrt{\pi}}\, \partial_x\Theta_{q} \right\} 
\\
\hat{\mathcal{K}}^b_{l,l+1}
 =&\sum_{q=1,2} \left[   -\frac{ a \,\text{Re}(J_z+\bar{J}_z)}{\pi \sqrt{\pi}}\, \partial_x\Phi_{q} 
  -\frac{a^2}{\pi} \text{Im}(J_z+\bar{J}_z) \partial_x\Theta_{q} \partial_x\Phi_{\bar{q}} \right. \nonumber \\
 &\left. \quad \quad -\frac{\text{Re}(J_z-\bar{J}_z)}{\pi^2} \mathrm{cos}\left(2\sqrt{\pi}\Phi_q\right)\mathrm{sin}\left(2\sqrt{\pi}\Phi_{\bar{q}}\right) \right]
\\
 \hat{\mathcal{K}}^c_{l,l+1}
 =&\sum_{q=1,2} \left[-\frac{ 2  a^2 \text{Re}(J_{z z})}{\pi^2} \left(\partial_x \Phi_q  \right)^2+
 \frac{2 \text{Re}(J_{z z})}{\pi^3}\cos\left(2\sqrt{\pi}\Phi_q\right)\cos\left(2\sqrt{\pi}\Phi_{\bar{q}}\right) 
 \right. \nonumber \\
 & \left. \quad \quad -\frac{\text{Re}(J_{z z})}{2 \pi^3}\cos\left(4\sqrt{\pi}\Phi_q\right)\right]\enspace.
 \end{align}
  We further introduce '+' and '-' - fields
\begin{align}
\Phi_{\pm}=\Phi_{1} \pm \Phi_{2} ,\quad \Theta_{\pm } = (\Theta_{1}\pm \Theta_{2})/2   
\end{align}
which yields
\begin{align}
 \hat{\mathcal{K}}^a_{l,l+1}   =& \frac{  a^2  \text{Re}(\tilde{J})}{2}\Big[2\left(\partial_x\Theta_{+}\right)^{2}+\frac{1}{2}\left(\partial_x\Phi_{+}\right)^{2} 
+2\left(\partial_x\Theta_{-}\right)^{2}+\frac{1}{2}\left(\partial_x\Phi_{-}\right)^{2} \Big] -\frac{ 2  a\, \text{Im}(\tilde{J})}{ \sqrt{\pi}}\, \partial_x\Theta_{+} 
\\
  \hat{\mathcal{K}}^b_{l,l+1}=   &-\frac{   a \,\text{Re}(J_z+\bar{J}_z)}{\pi \sqrt{\pi}}\, \partial_x\Phi_{+}-\frac{  a^2\, \text{Im}(J_z+\bar{J}_z)}{ \pi}\, \left[\partial_x\Theta_{+} \partial_x\Phi_{+}+\partial_x\Theta_{-} \partial_x\Phi_{-} \right] \nonumber \\
  & +\frac{\text{Re}(J_z-\bar{J}_z)}{\pi^2} \sin \left(2\sqrt{\pi}\Phi_+\right) \\
 \hat{\mathcal{K}}^c_{l,l+1}= &  -\frac{   a^2 \text{Re}(J_{z z})}{\pi^2}\left[\left(\partial_x \Phi_{+}\right)^2 + \left( \partial_x \Phi_{-}\right)^2\right] \nonumber \\
 & +\frac{2 \text{Re}(J_{z z})}{\pi^3}\left[ \cos\left(2\sqrt{\pi}\Phi_+\right) 
 +\cos\left(2\sqrt{\pi}\Phi_{-}\right) -\frac{1}{2}\cos\left(2\sqrt{\pi}\Phi_+\right)\cos\left(2\sqrt{\pi}\Phi_{-}\right)\right]\enspace.
\end{align}
Here, we omitted oscillating terms and contributions of higher order. The additional constant in Eq.~(\ref{nnBosonized}) entering $:\hat{n}_{l,q}:\, :\hat{n}_{l+1,q}:$  can be absorbed into the hopping amplitude of $\hat{\mathcal{K}}^a_{l,l+1}$, i.e. $\tilde{J}\to \tilde{J}-J_{zz}/2 \pi^2$.  Altogether, the intra-chain hopping  is given by  
\begin{align}
\hat{\cal K}^{\text{intra}}_{l,l+1}\simeq &  \frac{  a^2  \text{Re}\left(\tilde{J}-\frac{J_{z z}}{2\pi^2}\right)}{2}\Big[2\left(\partial_x\Theta_{+}\right)^{2}+\frac{1}{2}\left(\partial_x\Phi_{+}\right)^{2} 
+2\left(\partial_x\Theta_{-}\right)^{2}+\frac{1}{2}\left(\partial_x\Phi_{-}\right)^{2} \Big] \nonumber\\
&-\frac{   a^2 \text{Re}(J_{z z})}{\pi^2}\left[\left(\partial_x \Phi_{+}\right)^2 + \left( \partial_x \Phi_{-}\right)^2\right]-\frac{   a \,\text{Re}(J_z+\bar{J}_z)}{\pi \sqrt{\pi}}\, \partial_x\Phi_{+}-\frac{ 2  a\, \text{Im}(\tilde{J}-\frac{J_{z z}}{2\pi^2})}{ \sqrt{\pi}}\, \partial_x\Theta_{+} \nonumber \\
& -\frac{  a^2\, \text{Im}(J_z+\bar{J}_z)}{ \pi}\, \left[\partial_x\Theta_{+} \partial_x\Phi_{+}+\partial_x\Theta_{-} \partial_x\Phi_{-} \right] \nonumber \\
&+\frac{2 \text{Re}(J_{z z})}{\pi^3}\left[ \cos\left(2\sqrt{\pi}\Phi_+\right)+\cos\left(2\sqrt{\pi}\Phi_{-}\right) -\frac{1}{2}\cos\left(2\sqrt{\pi}\Phi_+\right)\cos\left(2\sqrt{\pi}\Phi_{-}\right)\right] \nonumber \\ &+\frac{\text{Re}(J_z-\bar{J}_z)}{\pi^2} \sin \left(2\sqrt{\pi}\Phi_+\right)\enspace.
\end{align}
The operator $\sin \left(2\sqrt{\pi}\Phi_+\right)$ is not allowed by symmetry as discussed in section \ref{symmetries}. Hence, the lattice model can only realize coupling constants for which this operator has zero amplitude. 
In the following we will therefore exclude this operator. 

The inter-chain hopping 
\begin{align}
    \hat{\cal K}^{\text{inter}}_{l,l+1}=-\frac{1}{2}\left({t^\star}\left(\hat{n}_{2,l},\hat{n}_{1,l+1}\right)\hat{c}^{\dagger}_{1,l}\hat{c}_{2,l+1}^{\phantom{\dagger}} + {t^\star}\left(\hat{n}_{1,l},\hat{n}_{2,l+1}\right)\hat{c}^{\dagger}_{2,l}\hat{c}_{1,l+1}^{\phantom{\dagger}}+ \rm{H.c.}\right)
\end{align}
    is  bosonized by returning to spin language and directly inserting the bosonized expressions for the spin operators, i.e. Eq.~(\ref{Bosonizationspin}), into the spin flip terms. This results in
\begin{align}
\hat{\cal K}^{\text{inter}}_{l,l+1}\simeq  -\frac{\text{Re}(\tilde{J}) }{\pi} \,\cos\left(2\sqrt{\pi}\Theta_{-}\right)\enspace
\end{align}
where we neglected further corrections caused by the gauge function which would only renormalize the prefactor in front of the $\cos\left(2\sqrt{\pi}\Theta_{-}\right)$-term but not generate further operator content.

\subsection{Bosonization of density-density interactions and complete Hamiltonian}
  Finally, we consider the onsite and nearest neighbor interaction terms 
  \begin{align}
    \hat{H}_{\rm int}=\sum_l\sum_{q=1,2}\left[\frac{U}{2} :\hat{n}_{q,l}::\hat{n}_{\bar{q},l}:+ V\left( :\hat{n}_{q,l}::\hat{n}_{q,l+1}:+:\hat{n}_{q,l}::\hat{n}_{\bar{q},l+1}:\right)\right]\enspace.
  \end{align} Bosonizing we obtain \cite{Giamarchi2003}
  \begin{align}
  :\hat{n}_{q,l}::\hat{n}_{q,l+1}:
  \simeq& -\frac{1}{2\pi^2} +\frac{ 2 a^2 }{\pi}\left(\partial_x \Phi_{q}\right)^2 
  + \frac{1}{2 \pi^2}\cos\left(4\sqrt{\pi}\Phi_{q}\right) \\
 : \hat{n}_{q,l}: :\hat{n}_{\bar{q},l^\prime}: \simeq & \frac{ a^2}{\pi}(\partial_x \Phi_{q} )(\partial_x \Phi_{\bar{q}}) \nonumber\\ 
  &+\frac{-1+2\delta_{l,l^\prime}}{2\pi^2} \Big[\cos \left(2\sqrt{\pi} (\Phi_{q}-\Phi_{\bar{q}})\right) -\cos \left(2\sqrt{\pi} (\Phi_{q}+\Phi_{\bar{q}})\right)\Big]
  \end{align}
  for $l^\prime=l,l+1$ where we have omitted oscillating contributions.
  Transforming into '+' and '-' - fields this yields
  \begin{align}
 \hat{H}_{\rm int} \simeq & \int dx \Big\{\frac{ a}{4 \pi} \left[(U+6V)\left(\partial_x \Phi_+\right)^2-(U-2V)\left(\partial_x \Phi_{-}\right)^2\right] \nonumber\\
 & +  \frac{U-2V}{2 \pi^2 a}\left[-\cos\left(2\sqrt{\pi}\Phi_+\right)+\cos\left(2\sqrt{\pi}\Phi_{-}\right)\right] 
 +\frac{V}{\pi^2 a}  \cos\left(2\sqrt{\pi}\Phi_+\right)\cos\left(2\sqrt{\pi}\Phi_{-}\right) \Big\} \enspace.
  \end{align}

We are now in the position to add up all terms and arrive at the Bosonized Hamiltonian 
\begin{align}
\hat{H}\simeq {\hat H}_{+} +{\hat H}_{-}+ {\hat H}_{+-}
\end{align} with
\begin{eqnarray}
{\hat H}_{+}&=&{\hat H}^{0}_{+} + \int dx \Big[ A(\partial_{x} \Theta_{+}) + B (\partial_{x} \Phi_{+})  + \Delta (\partial_{x} \Theta_{+})
(\partial_{x} \Phi_{+}) + \frac{g_{1}}{(\pi a)^{2}} \cos (2\sqrt{\pi}\Phi_{+})  \Big],\\
{\hat H}_{-}&=&{\hat H}^{0}_{-} + \int dx \Big[ \Delta (\partial_{x} \Theta_{-}) (\partial_{x} \Phi_{-}) + \frac{g_{2}}{(\pi a)^{2}} \cos (2\sqrt{\pi}\Phi_{-}) \nonumber
+ \frac{g_{3}}{(\pi a)^{2}} \cos(2\sqrt{\pi}\Theta_{-}) \Big], \\
{\hat H}_{+-}&=&  \frac{g_{4}}{(\pi a)^{2}} \int dx \cos(2\sqrt{\pi}\Phi_{+}) \cos(2\sqrt{\pi}\Phi_{-})  ,
\end{eqnarray}
where $\hat{H}^{0}_{\nu}$ for  $\nu=+,-$ denotes the ordinary Tomonaga-Luttinger Hamiltonian 
\begin{align}
\hat{H}^0_\nu=&\frac{u_\nu}{2 \pi}\int dx \left[K_\nu\left(\partial_x\Theta_\nu\right)^{2}+\frac{1}{K_\nu}\left(\partial_x\Phi_\nu\right)^{2}\right] \enspace.
\end{align}
We obtain for the Luttinger parameter to lowest order in the interactions
\begin{align}
K_{+}&=2\left(1+\frac{U+6V-4\,\text{Re}(J_{z z})/\pi}{\pi\text{Re}(\tilde{J}-J_{z z}/2\pi^2)}\right)^{-1/2},\\
K_{-}&=2\left(1-\frac{U-2V+4\,\text{Re}(J_{z z})/\pi}{\pi\text{Re}(\tilde{J}-J_{z z}/2\pi^2)}\right)^{-1/2}
\end{align}
while the velocities for each channel $\nu=+,-$ are given by
\begin{eqnarray}
u_{\nu}=a\,2\pi\text{Re}(\tilde{J}-J_{z z}/2\pi^2)/K_{\nu}\enspace.
\label{velocities}
\end{eqnarray}
Due to the asymmetry with respect to the Tomonaga-Luttinger basis we get linear couplings in the theory, i.e.
\begin{align}
A=& -\frac{ 2   }{ \sqrt{\pi}}\text{Im}\left(\tilde{J}-\frac{J_{z z}}{2\pi^2}\right)\label{linearcouplingphasethetanull}\\
B=&  -\frac{  1 }{\pi \sqrt{\pi}}\text{Re}(J_z+\bar{J}_z)\label{linearcoulingdenitythetanull}\\
\Delta= & -\frac{ a }{ \pi}\text{Im}(J_z+\bar{J}_z)\label{bilinearcouplings}
\end{align}
which can be interpreted as chemical potentials for the density and phase-excitations for each sector as well as couplings proportional to the momentum operator of each liquid. The latter implies that we have chosen a reference frame, where the system is not at rest.

%

The pinning terms that appear in our bosonized theory are given by the following coupling constants in the weak coupling limit
\begin{align}
&\frac{g_{1}}{a}=\frac{1}{2}(2V-U)+\frac{2 }{\pi}\text{Re}(J_{z z})\label{gonethetanull},
\\
&\frac{g_{2}}{a}=\frac{1}{2}(U-2V)+\frac{2 }{\pi}\text{Re}(J_{z z})\label{gtwothetanull},
\\
&\frac{g_{3}}{a}=- \pi \text{Re}(\tilde{J})\label{gthreethetanull},
\\
&\frac{g_4}{a}=V-\frac{\mathrm{Re}(J_{z z})}{\pi}\label{pinning}\enspace.
\end{align}
It must be noted here that the mapping from bosons to spin operators already implies a finite interaction in terms of $J_{zz}$, 
which is non-zero even for $\theta=U=V=0$.  Therefore, the weak-coupling limit is never exact and the actual coupling constant will
quantitatively differ from the formulas given above.  Nonetheless, the operator content and the qualitative behavior with increasing 
$\theta$, $V$ and $U$ is robust.

\begin{table}[t!]
	\centering
	\begin{tabular}{||l|l|l||} 
		\hline
		Symmetry & Lattice & Bosonic fields   \\ [0.5ex] 
		\hline\hline
		
		Translation & $\hat{b}_j\rightarrow\hat{b}_{j+1}$ & $\hat{\Phi}_+(x)\rightarrow\hat{\Phi}_+(x)+\pi$,$\phantom{a}\hat{\Theta}_+(x)\rightarrow\hat{\Theta}_+(x)$  \\ 
  & &$\hat{\Phi}_-(x)\rightarrow\hat{\Phi}_-(x)$,$\phantom{a}\hat{\Theta}_-(x)\rightarrow\hat{\Theta}_-(x)$ \\
  \hline
		Time reversal $\mathcal{K}$ & $\hat{b}_j\rightarrow\hat{b}_{j}$ & $\hat{\Phi}_+(x)\rightarrow\hat{\Phi}_+(x)$, $\phantom{a}\hat{\Theta}_+(x)\rightarrow-\hat{\Theta}_+(x)$  \\ 
  & & $\hat{\Phi}_-(x)\rightarrow\hat{\Phi}_-(x)$, $\phantom{a}\hat{\Theta}_-(x)\rightarrow-\hat{\Theta}_-(x)$ \\
  \hline
		$U(1)$ gauge  & $\hat{b}_j\rightarrow\hat{b}_{j}e^{i\alpha}$ &$\hat{\Phi}_+(x)\rightarrow\hat{\Phi}_+(x)$, $\phantom{a}\hat{\Theta}_+(x)\rightarrow\hat{\Theta}_+(x)+\alpha$  \\ 
  & &$\hat{\Phi}_-\rightarrow\hat{\Phi}_-(x)$, $\phantom{a}\hat{\Theta}_-(x)\rightarrow\hat{\Theta}_-(x)$ \\
  \hline
		Site parity $\mathcal{I}_-$ & $\hat{b}_j\rightarrow\hat{b}_{-j}$ & $\hat{\Phi}_+(x)\rightarrow-\hat{\Phi}_+(-x)+\pi$,    $\phantom{a}\hat{\Theta}_+(x)\rightarrow\hat{\Theta}_+(-x)$  \\
  &  & $\hat{\Phi}_-(x)\rightarrow-\hat{\Phi}_-(-x)$, $\phantom{a}\hat{\Theta}_-(x)\rightarrow\hat{\Theta}_-(-x)$\\
  \hline
		Link parity $\mathcal{I}$ & $\hat{b}_j\rightarrow\hat{b}_{1-j}$ & $\hat{\Phi}_+(x)\rightarrow-\hat{\Phi}_+(-x)$,  
		  $\phantom{a}\hat{\Theta}_+(x)\rightarrow\hat{\Theta}_+(-x)$ \\
    & & $\hat{\Phi}_-(x)\rightarrow-\hat{\Phi}_-(-x)$, $\phantom{a}\hat{\Theta}_-(x)\rightarrow\hat{\Theta}_-(-x)$
  \\     [0.5ex] 
		\hline
	\end{tabular}
	\caption{Symmetry operations and their realizations in Bosonic fields. }
	\label{tablesymmetries}
\end{table}

\section{Symmetries} \label{symmetries}
\label{appendC}

Here we discuss how the symmetry transformations on the lattice Hamiltonian are reflected by symmetry operations on the fields in the bosonization as summarized in Table \ref{tablesymmetries}.  This will be used to identify which operators are allowed or forbidden in the effective field theory Hamiltonian.

 The anyonic lattice model of interest $\hat{H}={\hat {H}_{\rm kin}}+ {\hat {H}_{\rm int}}$ with
\begin{align}
\label{kinetictermBoson}
{\hat {H}_{\rm kin}}&=-J{\sum_{l}}\left[{\hat b}^{\dagger}_{l}{\hat b}_{l+1} e^{i\theta {\hat n}_{l}}+{\rm H.c.}\right] \\
{\hat {H}_{\rm int}}&= \sum_{l}\left[ {\frac{U}{2}}{\hat n}_{l} ({\hat n}_{l}-1) + V  {\hat n}_{l} {\hat n}_{l+1} \right]
\end{align}
exhibits translational invariance and conservation of total particle number ($U(1)$ gauge). Link reflection symmetry $\mathcal{I}$ and time-reversal $\mathcal{K}$, in contrast, are broken. All symmetries are inherited by the
Bosonic low-energy description and Table~\ref{tablesymmetries} summarizes their
realizations.

A modified inversion symmetry is still
obeyed \cite{Lange2017}. The lattice model is invariant under the combined symmetry operation $\tilde{\mathcal{I}}=\hat{U}(\theta)\mathcal{K} \mathcal{I}$ where the non-linear unitary transformation $\hat{U}(\theta)=e^{-i \theta \sum_l \hat{n}_l (\hat{n}_l-1)/2 }$  generates a density-dependent  phase for  Bosonic operators \cite{Lange2017}
\begin{align}
{\hat b}^{\dagger}_{l} \to \hat{U}(\theta) {\hat b}^{\dagger}_{l} \hat{U}^\dagger(\theta)=\hat{b}^{\dagger}_{l} e^{-i \theta \hat{n}_l } \enspace.
\label{actionU}
\end{align}
 While the realizations of $\mathcal{K}$ and $ \mathcal{I}$ are known, see Table~\ref{tablesymmetries}, it is difficult to determine how the transformation $\hat{U}(\theta)$ acts on the Bosonic fields.   This is due to the fact that the density dependent phase
 $\exp (-i \theta \hat{n}_l)$ is represented by integer values of $n_l$ on the lattice, which ensures a $\theta\to \theta+2\pi$ periodicity.
 In the continuum limit there is no such lattice restriction, so it is far from trivial to implement shifts of the fields that have this topological property.  From Eq.~(\ref{actionU}) and section
\ref{sec:Peierls} we know that $\hat{U}(\theta)$ generates factors of the form as in Eq.~(\ref{exp_sz}) above which can be absorbed by redefinition of the coefficients $a(\theta)$ and $b(\theta)$ which in turn determine the coupling constants $\tilde{J},J_z,\bar{J}_z$ and $J_{zz}$ of the spin-1/2 two-leg ladder model. Such non-linear transformations cannot be tracked in the bosonized expressions of $\hat{S}^{\pm}_{q,l}$ in Eq.~(\ref{Bosonizationspin}) and therefore the realization of the combined symmetry operation $\hat{U}(\theta)\mathcal{K} \mathcal{I}$ in $\Theta_\pm$ is unclear.
 
 On the other hand, it is clear that densities $\hat{b}^{\dagger}_{l} \hat{b}^{\phantom{\dagger}}_{l} =\hat{n}_l=-\hat{S}^z_{1,l}-\hat{S}^z_{2,l}+1$ are not affected by  $\hat{U}(\theta)$, so therefore
any transformation on the fields due to $\hat{U}(\theta)$ must also leave the spin-z operators 
 in Eq.~(\ref{BosonizationspinSz}) invariant.  Therefore, shifts on the $\Phi$-fields cannot be generated by $\hat{U}(\theta)$, while 
 we cannot make any statement how the $\Theta$-fields are transformed.

Next, we turn to the modified inversion symmetry which is represented by the  combined symmetry operation $\tilde{\mathcal{I}}=\hat{U}(\theta)\mathcal{K} \mathcal{I}$.  For densities in Eq.~(\ref{BosonizationspinSz}) this operation corresponds to a simple inversion symmetry 
$S_{q,l}^z\to S_{q,1-l}^z$, since $\mathcal K$ and $\hat{U}(\theta)$ do not affect the densities.
  We therefore conclude that the Bosonic $\Phi$-fields transform as  
\begin{align}
    \Phi_q(x)\to -\Phi_q(-x) \quad q=1,2
\end{align} 
and
\begin{align}
    \Phi_+(x)\to -\Phi_+(-x), \quad \Phi_-(x)\to -\Phi_-(-x)
    \label{trans_phi}
\end{align} 
under the action of the combined symmetry $\hat{U}(\theta)\mathcal{K} \mathcal{I}$. 
Even without knowing the symmetry action on the $\Theta$-fields,  
the transformation property of $\Phi_+(x)$ in Eq.~(\ref{trans_phi}) rules out the existence of the $\sin(\sqrt{4\pi} \Phi_{+}(x))$ - operator as a perturbation. Note, that the operators $\Delta \partial_{x} \Theta_{\pm}(x)
\partial_{x} \Phi_{\pm}(x)$ appearing in the bosonized anyonic Hamiltonian in Eq.~(7) of the  main text obey the combined $\mathcal{K} \mathcal{I}$ symmetry, i.e. $\hat{\Phi}_\pm(x)\rightarrow-\hat{\Phi}_\pm(-x)$ and $\hat{\Theta}_\pm(x)\rightarrow-\hat{\Theta}_\pm(-x)$, see Table~\ref{tablesymmetries}.

\section{Additional numerical data for order parameters, fidelity suscpetibility, entanglement entropy, and correlation lengths }
\label{AppendixNumerics} 
\label{appendD}

In Figs.~\ref{Paper1}-\ref{Paper4} additional numerical data for the order parameters, the fidelity susceptibility, the entanglement entropy and the correlation length across phase transition lines is presented to illustrate how the phase transition lines were determined and analyzed.  In addition to the order parameters as shown in Fig.~\ref{Paper1}, we used the fidelity susceptibility in Figs.~\ref{Crossover} and \ref{Crossovertheta0p25} and the entangle entropy in Figs.~\ref{Paper3} amd \ref{Paper4}.  We will define and explain those quantities in the
following.

\begin{figure}[b!]
  \includegraphics[width=0.75\textwidth]{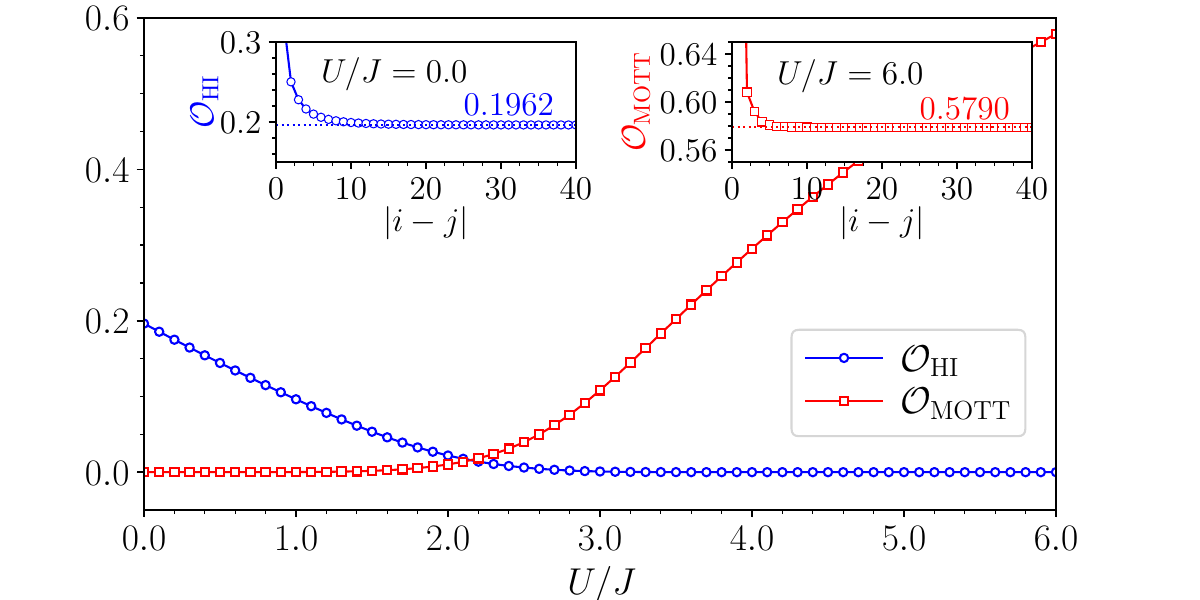}
  \vspace{-0.3cm}
  \caption{String order parameters $\mathcal{O}_{\mathrm{HI}}$  and $\mathcal{O}_\mathrm{Mott}$  plotted along the HI to Mott transition as a function of $U/J$, showing the existing hidden order in both phases at $V/J=1.0, \theta=0$. The inlets show how the values were obtained at example points deep in the phases at $U/J=0.0$ (HI phase) and $U/J=6.0$ (Mott phase) respectively. $\mathcal{O}(|i-j|)$ is extrapolated to $|i-j| \rightarrow \infty$ marked by the dotted lines. Data shown from finite DMRG, $L=100, M=400$, periodic boundary conditions.}
  \label{Paper1}
\end{figure}
\begin{figure}[h!]
  \includegraphics[width=0.85\textwidth]{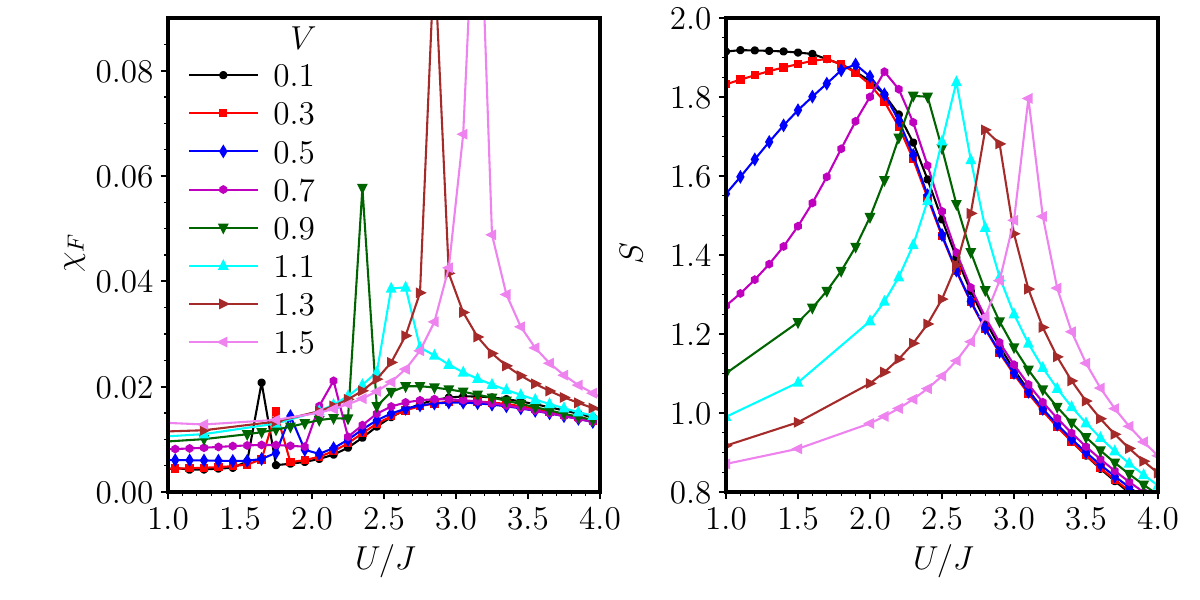}
  \vspace{-0.3cm}
  \caption{Fidelity susceptibility $\chi_\mathrm{F}$ (left) and entanglement entropy $S$ (right) plotted over $U/J$ across the Mott to HI transition for $\theta=0$  and various fixed values of $V/J$. Data shown from iDMRG, $M=400$.}
  \label{Crossover}
\end{figure}

\begin{figure}[h!]
	\includegraphics[width=0.6\textwidth]{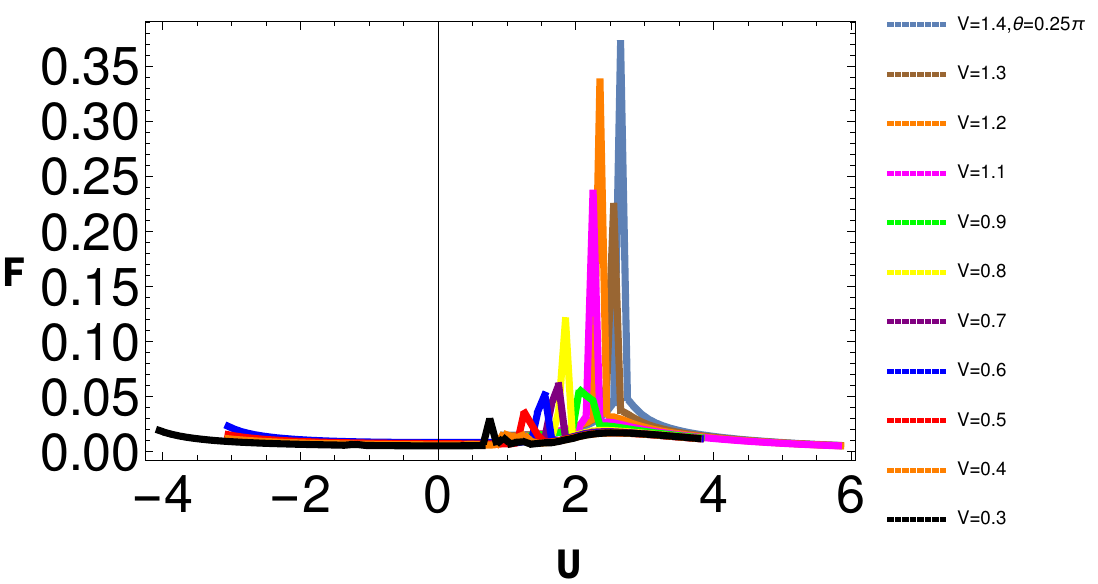}
	\vspace{-0.3cm}
	\caption{Fidelity susceptibility $\chi_\mathrm{F}$ for $\theta=0.25$ plotted over $U/J$ across the Mott to HI transition for various fixed values of $V/J$. Data shown from iDMRG, $M=400$.}
	\label{Crossovertheta0p25}
\end{figure}

\begin{figure}[h!]
	\includegraphics[width=0.7\textwidth]{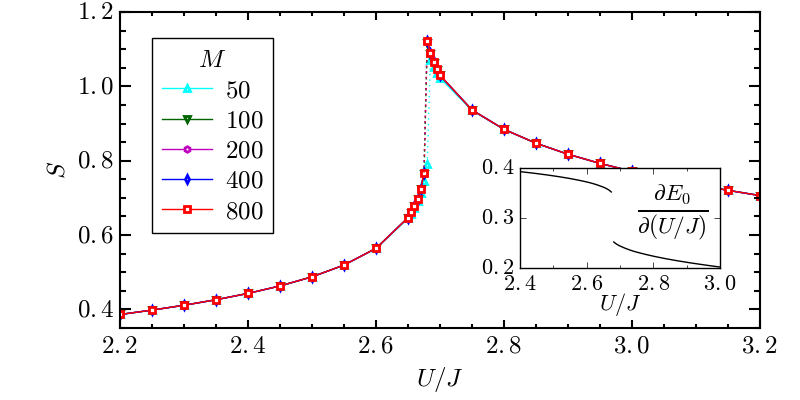}
	\vspace{-0.7cm}
	\caption{Entanglement entropy $S$ (large plot) and partial derivative of ground-state energy with $U/J$ (inlet) plotted over $U/J$ at fixed $V/J=3.0, \theta=\pi$ across the CDW to Mott transition line. Both quantities show a jump characteristic of a 1st order transition. Data from iDMRG, $M=50$ to $M=800$ to exclude the possiblity of an accuracy error. }
	\label{Paper3}
\end{figure}
\begin{figure}[h!]
	\includegraphics[width=0.8\textwidth]{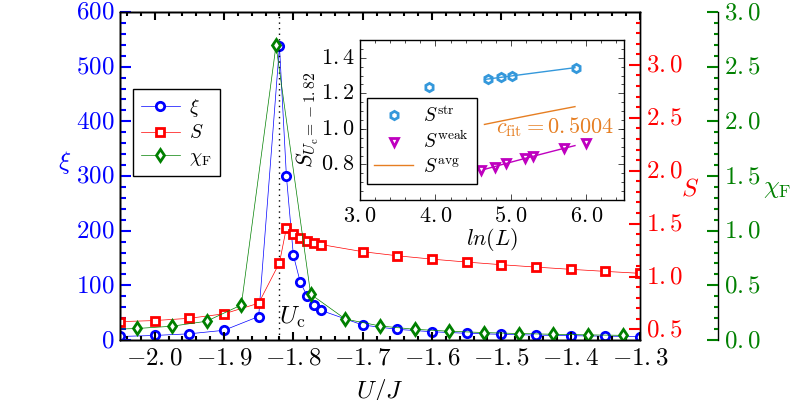}
	\vspace{0.3cm}
	\caption{Correlation length $\xi$, entanglement entropy $S$ and fidelity susceptibility $\chi_\mathrm{F}$ plotted over $U/J$ along the dimer to Mott transition line at $V/J=0.2, \theta=\pi$ showing a 2nd order transition at $U_c/J=-1.82$.  Data from iDMRG, $M=800$. Inlet shows entanglement entropy at the transition point $V/J=0.2, U_\mathrm{c}/J=-1.82$ plotted over $\ln(L)$, the slope of which is given by $c/6$. Depending on the cut being along a strong or weak bond, we get $S^{\mathrm{str}}$ and $S^{\mathrm{weak}}$. The slope of the averaged entropy $S^{\mathrm{avg}}$ corresponds to a central charge $c=0.5004$. Inlet data from DMRG with $M=400$ using open boundary conditions, $S$ measured on middle bond for minimal boundary effect.}
	\label{Paper4}
\end{figure}

A special property of one-dimensional, gapped Hamiltonians  can be conveniently formalized by means of the entanglement entropy. The quantity is defined as the Von Neumann entropy, i.e.
\begin{align}
	\label{Vonneumannentropy}
	S(\hat{\rho})=-\mathrm{tr}\hat{\rho}\ln(\hat{\rho})
\end{align}   
of the so-called reduced density matrix $\hat{\rho}\equiv\hat{\rho}_{R,L}$, obtainable by partially tracing out a sub-system from the pure density matrix,
\begin{align}
	\label{reduceddensitymatricx}
	\hat{\rho}_{R,L}=\mathrm{tr}_{L,R}\hat{\rho}.
\end{align}  
The entanglement entropy in Eq.~(\ref{Vonneumannentropy}) is thereby equivalent for each $L,R$ following from the Schmidt decomposition of an arbitrary pure state into two orthonormal bases. $S(\rho_{L,R})\geq 0$ quantifies how entangled the two subsystems are, with $S(\rho_{L,R})=0$ denoting the special case that both sub-systems are in product form, and fully unentangled. 

The von Neumann entanglement entropy diverges logarithmically in the correlation length $\xi$ when approaching a critical point  \cite{Nishimoto2011}, i.e.
\begin{align}
	\label{centralcharge}
	S(\rho_{L,R})\propto c\ln(\xi) 
\end{align}  
with $c$ being the central charge of the underlying two-dimensional conformal field theory. For finite size systems the correlation length at the continuous transition is limited by the system size $L$.  At
the critical point, the entanglement entropy  then diverges as 
\begin{align}
	\label{centralcharge}
	S(\rho_{L,R})  = \frac{c}{6}\ln(L) 
\end{align} 
where the factor $1/6$ applies to open boundary conditions  \cite{Nishimoto2011}. 

Another useful quantity to observe the occurrence of phase transitions in the framework of DMRG  is the fidelity of quantum states $F$ and especially its susceptibility $\chi$. The fidelity is defined as the absolute value of the overlap between a quantum state $|{\Psi(\lambda)}\rangle$ depending on a parameter $\lambda$ with a state with a slightly shifted parameter
$|\Psi(\lambda+\delta\lambda)\rangle$,i.e.
\begin{align}
	\label{Fidelity}
	F=\vert \langle{\Psi(\lambda)}|{\Psi(\lambda+\delta\lambda)}\rangle\vert.
\end{align}
It is convenient to define its logarithmic derivative, 
\begin{align}
	\label{Fidelitysusceptibility}
	\chi_F=-\frac{\partial^2\ln(F)}{\partial\delta\lambda^2}\vert_{\delta\lambda=0}.
\end{align} 
Here, the parameter $\lambda$ is the free parameter along which line the phase transition is to be located. 

\newpage

\end{widetext}
\bibliography{AnyonPaperBibliography.bib}

\begin{thebibliography}{57}%
\makeatletter
\providecommand \@ifxundefined [1]{%
 \@ifx{#1\undefined}
}%
\providecommand \@ifnum [1]{%
 \ifnum #1\expandafter \@firstoftwo
 \else \expandafter \@secondoftwo
 \fi
}%
\providecommand \@ifx [1]{%
 \ifx #1\expandafter \@firstoftwo
 \else \expandafter \@secondoftwo
 \fi
}%
\providecommand \natexlab [1]{#1}%
\providecommand \enquote  [1]{``#1''}%
\providecommand \bibnamefont  [1]{#1}%
\providecommand \bibfnamefont [1]{#1}%
\providecommand \citenamefont [1]{#1}%
\providecommand \href@noop [0]{\@secondoftwo}%
\providecommand \href [0]{\begingroup \@sanitize@url \@href}%
\providecommand \@href[1]{\@@startlink{#1}\@@href}%
\providecommand \@@href[1]{\endgroup#1\@@endlink}%
\providecommand \@sanitize@url [0]{\catcode `\\12\catcode `\$12\catcode
  `\&12\catcode `\#12\catcode `\^12\catcode `\_12\catcode `\%12\relax}%
\providecommand \@@startlink[1]{}%
\providecommand \@@endlink[0]{}%
\providecommand \url  [0]{\begingroup\@sanitize@url \@url }%
\providecommand \@url [1]{\endgroup\@href {#1}{\urlprefix }}%
\providecommand \urlprefix  [0]{URL }%
\providecommand \Eprint [0]{\href }%
\providecommand \doibase [0]{https://doi.org/}%
\providecommand \selectlanguage [0]{\@gobble}%
\providecommand \bibinfo  [0]{\@secondoftwo}%
\providecommand \bibfield  [0]{\@secondoftwo}%
\providecommand \translation [1]{[#1]}%
\providecommand \BibitemOpen [0]{}%
\providecommand \bibitemStop [0]{}%
\providecommand \bibitemNoStop [0]{.\EOS\space}%
\providecommand \EOS [0]{\spacefactor3000\relax}%
\providecommand \BibitemShut  [1]{\csname bibitem#1\endcsname}%
\let\auto@bib@innerbib\@empty
\bibitem [{\citenamefont {Leinaas}\ and\ \citenamefont
  {Myrheim}(1977)}]{anyons}%
  \BibitemOpen
  \bibfield  {author} {\bibinfo {author} {\bibfnamefont {J.}~\bibnamefont
  {Leinaas}}\ and\ \bibinfo {author} {\bibfnamefont {J.}~\bibnamefont
  {Myrheim}},\ }\bibfield  {title} {\bibinfo {title} {On the theory of
  identical particles},\ }\href {https://doi.org/10.1007/BF02727953} {\bibfield
   {journal} {\bibinfo  {journal} {Nuovo Cimento Soc. Ital. Fis.}\ }\textbf
  {\bibinfo {volume} {37B}},\ \bibinfo {pages} {1} (\bibinfo {year}
  {1977})}\BibitemShut {NoStop}%
\bibitem [{foo()}]{footnote}%
  \BibitemOpen
  \href@noop {} {\bibinfo {title} {In the original paper \cite{anyons}
  fractional statistics for both two-dimensional and one-dimensional geometries
  are discussed. {O}ur work refers to the 1{D} case.}}\BibitemShut {Stop}%
\bibitem [{\citenamefont {Eckardt}(2017)}]{Eckardt2017}%
  \BibitemOpen
  \bibfield  {author} {\bibinfo {author} {\bibfnamefont {A.}~\bibnamefont
  {Eckardt}},\ }\bibfield  {title} {\bibinfo {title} {Colloquium: Atomic
  quantum gases in periodically driven optical lattices},\ }\href
  {https://doi.org/10.1103/RevModPhys.89.011004} {\bibfield  {journal}
  {\bibinfo  {journal} {Rev. Mod. Phys.}\ }\textbf {\bibinfo {volume} {89}},\
  \bibinfo {pages} {011004} (\bibinfo {year} {2017})}\BibitemShut {NoStop}%
\bibitem [{\citenamefont {Goldman}\ and\ \citenamefont
  {Dalibard}(2014)}]{Goldman2014}%
  \BibitemOpen
  \bibfield  {author} {\bibinfo {author} {\bibfnamefont {N.}~\bibnamefont
  {Goldman}}\ and\ \bibinfo {author} {\bibfnamefont {J.}~\bibnamefont
  {Dalibard}},\ }\bibfield  {title} {\bibinfo {title} {Periodically driven
  quantum systems: Effective {H}amiltonians and engineered gauge fields},\
  }\href {https://doi.org/10.1103/PhysRevX.4.031027} {\bibfield  {journal}
  {\bibinfo  {journal} {Phys. Rev. X}\ }\textbf {\bibinfo {volume} {4}},\
  \bibinfo {pages} {031027} (\bibinfo {year} {2014})}\BibitemShut {NoStop}%
\bibitem [{\citenamefont {Bukov}\ \emph {et~al.}(2015)\citenamefont {Bukov},
  \citenamefont {D'Alessio},\ and\ \citenamefont {Polkovnikov}}]{Bukov2015}%
  \BibitemOpen
  \bibfield  {author} {\bibinfo {author} {\bibfnamefont {M.}~\bibnamefont
  {Bukov}}, \bibinfo {author} {\bibfnamefont {L.}~\bibnamefont {D'Alessio}},\
  and\ \bibinfo {author} {\bibfnamefont {A.}~\bibnamefont {Polkovnikov}},\
  }\bibfield  {title} {\bibinfo {title} {Universal high-frequency behavior of
  periodically driven systems: from dynamical stabilization to {F}loquet
  engineering},\ }\href {https://doi.org/10.1080/00018732.2015.1055918}
  {\bibfield  {journal} {\bibinfo  {journal} {Advances in Physics}\ }\textbf
  {\bibinfo {volume} {64}},\ \bibinfo {pages} {139} (\bibinfo {year}
  {2015})}\BibitemShut {NoStop}%
\bibitem [{\citenamefont {Keilmann}\ \emph {et~al.}(2011)\citenamefont
  {Keilmann}, \citenamefont {Lanzmich}, \citenamefont {McCulloch},\ and\
  \citenamefont {Roncaglia}}]{Keilmann2011}%
  \BibitemOpen
  \bibfield  {author} {\bibinfo {author} {\bibfnamefont {T.}~\bibnamefont
  {Keilmann}}, \bibinfo {author} {\bibfnamefont {S.}~\bibnamefont {Lanzmich}},
  \bibinfo {author} {\bibfnamefont {I.}~\bibnamefont {McCulloch}},\ and\
  \bibinfo {author} {\bibfnamefont {M.}~\bibnamefont {Roncaglia}},\ }\bibfield
  {title} {\bibinfo {title} {Statistically induced phase transitions and anyons
  in 1{D} optical lattices},\ }\href {https://doi.org/10.1038/ncomms1353}
  {\bibfield  {journal} {\bibinfo  {journal} {Nature Communications}\ }\textbf
  {\bibinfo {volume} {2}},\ \bibinfo {pages} {361 EP } (\bibinfo {year}
  {2011})},\ \bibinfo {note} {article}\BibitemShut {NoStop}%
\bibitem [{\citenamefont {Greschner}\ and\ \citenamefont
  {Santos}(2015)}]{Greschner2015}%
  \BibitemOpen
  \bibfield  {author} {\bibinfo {author} {\bibfnamefont {S.}~\bibnamefont
  {Greschner}}\ and\ \bibinfo {author} {\bibfnamefont {L.}~\bibnamefont
  {Santos}},\ }\bibfield  {title} {\bibinfo {title} {Anyon {H}ubbard model in
  one-dimensional optical lattices},\ }\href
  {https://doi.org/10.1103/PhysRevLett.115.053002} {\bibfield  {journal}
  {\bibinfo  {journal} {Phys. Rev. Lett.}\ }\textbf {\bibinfo {volume} {115}},\
  \bibinfo {pages} {053002} (\bibinfo {year} {2015})}\BibitemShut {NoStop}%
\bibitem [{\citenamefont {Lange}\ \emph
  {et~al.}(2017{\natexlab{a}})\citenamefont {Lange}, \citenamefont {Ejima},\
  and\ \citenamefont {Fehske}}]{Lange2017}%
  \BibitemOpen
  \bibfield  {author} {\bibinfo {author} {\bibfnamefont {F.}~\bibnamefont
  {Lange}}, \bibinfo {author} {\bibfnamefont {S.}~\bibnamefont {Ejima}},\ and\
  \bibinfo {author} {\bibfnamefont {H.}~\bibnamefont {Fehske}},\ }\bibfield
  {title} {\bibinfo {title} {Anyonic {H}aldane insulator in one dimension},\
  }\href {https://doi.org/10.1103/PhysRevLett.118.120401} {\bibfield  {journal}
  {\bibinfo  {journal} {Phys. Rev. Lett.}\ }\textbf {\bibinfo {volume} {118}},\
  \bibinfo {pages} {120401} (\bibinfo {year} {2017}{\natexlab{a}})}\BibitemShut
  {NoStop}%
\bibitem [{\citenamefont {Str\"ater}\ \emph {et~al.}(2016)\citenamefont
  {Str\"ater}, \citenamefont {Srivastava},\ and\ \citenamefont
  {Eckardt}}]{Straeter2016}%
  \BibitemOpen
  \bibfield  {author} {\bibinfo {author} {\bibfnamefont {C.}~\bibnamefont
  {Str\"ater}}, \bibinfo {author} {\bibfnamefont {S.~C.~L.}\ \bibnamefont
  {Srivastava}},\ and\ \bibinfo {author} {\bibfnamefont {A.}~\bibnamefont
  {Eckardt}},\ }\bibfield  {title} {\bibinfo {title} {{F}loquet realization and
  signatures of one-dimensional anyons in an optical lattice},\ }\href
  {https://doi.org/10.1103/PhysRevLett.117.205303} {\bibfield  {journal}
  {\bibinfo  {journal} {Phys. Rev. Lett.}\ }\textbf {\bibinfo {volume} {117}},\
  \bibinfo {pages} {205303} (\bibinfo {year} {2016})}\BibitemShut {NoStop}%
\bibitem [{\citenamefont {Tang}\ \emph {et~al.}(2015)\citenamefont {Tang},
  \citenamefont {Eggert},\ and\ \citenamefont {Pelster}}]{Tang2015}%
  \BibitemOpen
  \bibfield  {author} {\bibinfo {author} {\bibfnamefont {G.}~\bibnamefont
  {Tang}}, \bibinfo {author} {\bibfnamefont {S.}~\bibnamefont {Eggert}},\ and\
  \bibinfo {author} {\bibfnamefont {A.}~\bibnamefont {Pelster}},\ }\bibfield
  {title} {\bibinfo {title} {Ground-state properties of anyons in a
  one-dimensional lattice},\ }\href
  {https://doi.org/10.1088/1367-2630/17/12/123016} {\bibfield  {journal}
  {\bibinfo  {journal} {New Journal of Physics}\ }\textbf {\bibinfo {volume}
  {17}},\ \bibinfo {pages} {123016} (\bibinfo {year} {2015})}\BibitemShut
  {NoStop}%
\bibitem [{\citenamefont {Lange}\ \emph
  {et~al.}(2017{\natexlab{b}})\citenamefont {Lange}, \citenamefont {Ejima},\
  and\ \citenamefont {Fehske}}]{Lange2017PRA}%
  \BibitemOpen
  \bibfield  {author} {\bibinfo {author} {\bibfnamefont {F.}~\bibnamefont
  {Lange}}, \bibinfo {author} {\bibfnamefont {S.}~\bibnamefont {Ejima}},\ and\
  \bibinfo {author} {\bibfnamefont {H.}~\bibnamefont {Fehske}},\ }\bibfield
  {title} {\bibinfo {title} {Strongly repulsive anyons in one dimension},\
  }\href {https://doi.org/10.1103/PhysRevA.95.063621} {\bibfield  {journal}
  {\bibinfo  {journal} {Phys. Rev. A}\ }\textbf {\bibinfo {volume} {95}},\
  \bibinfo {pages} {063621} (\bibinfo {year} {2017}{\natexlab{b}})}\BibitemShut
  {NoStop}%
\bibitem [{\citenamefont {Liu}\ \emph {et~al.}(2018)\citenamefont {Liu},
  \citenamefont {Garrison}, \citenamefont {Deng}, \citenamefont {Gong},\ and\
  \citenamefont {Gorshkov}}]{Liu2018a}%
  \BibitemOpen
  \bibfield  {author} {\bibinfo {author} {\bibfnamefont {F.}~\bibnamefont
  {Liu}}, \bibinfo {author} {\bibfnamefont {J.~R.}\ \bibnamefont {Garrison}},
  \bibinfo {author} {\bibfnamefont {D.-L.}\ \bibnamefont {Deng}}, \bibinfo
  {author} {\bibfnamefont {Z.-X.}\ \bibnamefont {Gong}},\ and\ \bibinfo
  {author} {\bibfnamefont {A.~V.}\ \bibnamefont {Gorshkov}},\ }\bibfield
  {title} {\bibinfo {title} {Asymmetric particle transport and light-cone
  dynamics induced by anyonic statistics},\ }\href
  {https://doi.org/10.1103/PhysRevLett.121.250404} {\bibfield  {journal}
  {\bibinfo  {journal} {Phys. Rev. Lett.}\ }\textbf {\bibinfo {volume} {121}},\
  \bibinfo {pages} {250404} (\bibinfo {year} {2018})}\BibitemShut {NoStop}%
\bibitem [{\citenamefont {Bonkhoff}\ \emph {et~al.}(2021)\citenamefont
  {Bonkhoff}, \citenamefont {J\"agering}, \citenamefont {Eggert}, \citenamefont
  {Pelster}, \citenamefont {Thorwart},\ and\ \citenamefont
  {Posske}}]{Bonkhoff2021}%
  \BibitemOpen
  \bibfield  {author} {\bibinfo {author} {\bibfnamefont {M.}~\bibnamefont
  {Bonkhoff}}, \bibinfo {author} {\bibfnamefont {K.}~\bibnamefont
  {J\"agering}}, \bibinfo {author} {\bibfnamefont {S.}~\bibnamefont {Eggert}},
  \bibinfo {author} {\bibfnamefont {A.}~\bibnamefont {Pelster}}, \bibinfo
  {author} {\bibfnamefont {M.}~\bibnamefont {Thorwart}},\ and\ \bibinfo
  {author} {\bibfnamefont {T.}~\bibnamefont {Posske}},\ }\bibfield  {title}
  {\bibinfo {title} {{B}osonic continuum theory of one-dimensional lattice
  anyons},\ }\href {https://doi.org/10.1103/PhysRevLett.126.163201} {\bibfield
  {journal} {\bibinfo  {journal} {Phys. Rev. Lett.}\ }\textbf {\bibinfo
  {volume} {126}},\ \bibinfo {pages} {163201} (\bibinfo {year}
  {2021})}\BibitemShut {NoStop}%
\bibitem [{\citenamefont {Bonkhoff}\ \emph {et~al.}(2023)\citenamefont
  {Bonkhoff}, \citenamefont {J\"ager}, \citenamefont {Schneider}, \citenamefont
  {Pelster},\ and\ \citenamefont {Eggert}}]{Bonkhoff2023}%
  \BibitemOpen
  \bibfield  {author} {\bibinfo {author} {\bibfnamefont {M.}~\bibnamefont
  {Bonkhoff}}, \bibinfo {author} {\bibfnamefont {S.~B.}\ \bibnamefont
  {J\"ager}}, \bibinfo {author} {\bibfnamefont {I.}~\bibnamefont {Schneider}},
  \bibinfo {author} {\bibfnamefont {A.}~\bibnamefont {Pelster}},\ and\ \bibinfo
  {author} {\bibfnamefont {S.}~\bibnamefont {Eggert}},\ }\bibfield  {title}
  {\bibinfo {title} {Coherence properties of the repulsive anyon-{H}ubbard
  dimer},\ }\href {https://doi.org/10.1103/PhysRevB.108.155134} {\bibfield
  {journal} {\bibinfo  {journal} {Phys. Rev. B}\ }\textbf {\bibinfo {volume}
  {108}},\ \bibinfo {pages} {155134} (\bibinfo {year} {2023})}\BibitemShut
  {NoStop}%
\bibitem [{\citenamefont {Görg}\ \emph {et~al.}(2019)\citenamefont {Görg},
  \citenamefont {Sandholzer}, \citenamefont {Minguzzi}, \citenamefont
  {Desbuquois}, \citenamefont {Messer},\ and\ \citenamefont
  {Esslinger}}]{Goerg2019}%
  \BibitemOpen
  \bibfield  {author} {\bibinfo {author} {\bibfnamefont {F.}~\bibnamefont
  {Görg}}, \bibinfo {author} {\bibfnamefont {K.}~\bibnamefont {Sandholzer}},
  \bibinfo {author} {\bibfnamefont {J.}~\bibnamefont {Minguzzi}}, \bibinfo
  {author} {\bibfnamefont {R.}~\bibnamefont {Desbuquois}}, \bibinfo {author}
  {\bibfnamefont {M.}~\bibnamefont {Messer}},\ and\ \bibinfo {author}
  {\bibfnamefont {T.}~\bibnamefont {Esslinger}},\ }\bibfield  {title} {\bibinfo
  {title} {Realization of density-dependent {P}eierls phases to engineer
  quantized gauge fields coupled to ultracold matter},\ }\href
  {https://doi.org/10.1038/s41567-019-0615-4} {\bibfield  {journal} {\bibinfo
  {journal} {Nat. Phys.}\ }\textbf {\bibinfo {volume} {15}},\ \bibinfo {pages}
  {1161} (\bibinfo {year} {2019})}\BibitemShut {NoStop}%
\bibitem [{\citenamefont {Lienhard}\ \emph {et~al.}(2020)\citenamefont
  {Lienhard}, \citenamefont {Scholl}, \citenamefont {Weber}, \citenamefont
  {Barredo}, \citenamefont {de~L\'es\'eleuc}, \citenamefont {Bai},
  \citenamefont {Lang}, \citenamefont {Fleischhauer}, \citenamefont
  {B\"uchler}, \citenamefont {Lahaye},\ and\ \citenamefont
  {Browaeys}}]{Lienhard2020}%
  \BibitemOpen
  \bibfield  {author} {\bibinfo {author} {\bibfnamefont {V.}~\bibnamefont
  {Lienhard}}, \bibinfo {author} {\bibfnamefont {P.}~\bibnamefont {Scholl}},
  \bibinfo {author} {\bibfnamefont {S.}~\bibnamefont {Weber}}, \bibinfo
  {author} {\bibfnamefont {D.}~\bibnamefont {Barredo}}, \bibinfo {author}
  {\bibfnamefont {S.}~\bibnamefont {de~L\'es\'eleuc}}, \bibinfo {author}
  {\bibfnamefont {R.}~\bibnamefont {Bai}}, \bibinfo {author} {\bibfnamefont
  {N.}~\bibnamefont {Lang}}, \bibinfo {author} {\bibfnamefont {M.}~\bibnamefont
  {Fleischhauer}}, \bibinfo {author} {\bibfnamefont {H.~P.}\ \bibnamefont
  {B\"uchler}}, \bibinfo {author} {\bibfnamefont {T.}~\bibnamefont {Lahaye}},\
  and\ \bibinfo {author} {\bibfnamefont {A.}~\bibnamefont {Browaeys}},\
  }\bibfield  {title} {\bibinfo {title} {Realization of a density-dependent
  {P}eierls phase in a synthetic, spin-orbit coupled {R}ydberg system},\ }\href
  {https://doi.org/10.1103/PhysRevX.10.021031} {\bibfield  {journal} {\bibinfo
  {journal} {Phys. Rev. X}\ }\textbf {\bibinfo {volume} {10}},\ \bibinfo
  {pages} {021031} (\bibinfo {year} {2020})}\BibitemShut {NoStop}%
\bibitem [{\citenamefont {Kwan}\ \emph {et~al.}()\citenamefont {Kwan},
  \citenamefont {Segura}, \citenamefont {Li}, \citenamefont {Kim},
  \citenamefont {Gorshkov}, \citenamefont {Eckardt}, \citenamefont
  {Bakkali-Hassani1},\ and\ \citenamefont {Greiner}}]{Kwan2023}%
  \BibitemOpen
  \bibfield  {author} {\bibinfo {author} {\bibfnamefont {J.}~\bibnamefont
  {Kwan}}, \bibinfo {author} {\bibfnamefont {P.}~\bibnamefont {Segura}},
  \bibinfo {author} {\bibfnamefont {Y.}~\bibnamefont {Li}}, \bibinfo {author}
  {\bibfnamefont {S.}~\bibnamefont {Kim}}, \bibinfo {author} {\bibfnamefont
  {A.~V.}\ \bibnamefont {Gorshkov}}, \bibinfo {author} {\bibfnamefont
  {A.}~\bibnamefont {Eckardt}}, \bibinfo {author} {\bibfnamefont
  {B.}~\bibnamefont {Bakkali-Hassani1}},\ and\ \bibinfo {author} {\bibfnamefont
  {M.}~\bibnamefont {Greiner}},\ }\bibfield  {title} {\bibinfo {title}
  {Realization of 1{D} anyons with arbitrary statistical phase},\ }\bibfield
  {journal} {\bibinfo  {journal} {arXiv:2306.01737}\ }\href
  {https://doi.org/https://doi.org/10.48550/arXiv.2306.01737}
  {https://doi.org/10.48550/arXiv.2306.01737}\BibitemShut {NoStop}%
\bibitem [{\citenamefont {Bloch}\ \emph {et~al.}(2008)\citenamefont {Bloch},
  \citenamefont {Dalibard},\ and\ \citenamefont {Zwerger}}]{Bloch2008}%
  \BibitemOpen
  \bibfield  {author} {\bibinfo {author} {\bibfnamefont {I.}~\bibnamefont
  {Bloch}}, \bibinfo {author} {\bibfnamefont {J.}~\bibnamefont {Dalibard}},\
  and\ \bibinfo {author} {\bibfnamefont {W.}~\bibnamefont {Zwerger}},\
  }\bibfield  {title} {\bibinfo {title} {Many-body physics with ultracold
  gases},\ }\href {https://doi.org/10.1103/RevModPhys.80.885} {\bibfield
  {journal} {\bibinfo  {journal} {Rev. Mod. Phys.}\ }\textbf {\bibinfo {volume}
  {80}},\ \bibinfo {pages} {885} (\bibinfo {year} {2008})}\BibitemShut
  {NoStop}%
\bibitem [{\citenamefont {Aidelsburger}\ \emph {et~al.}(2013)\citenamefont
  {Aidelsburger}, \citenamefont {Atala}, \citenamefont {Lohse}, \citenamefont
  {Barreiro}, \citenamefont {Paredes},\ and\ \citenamefont
  {Bloch}}]{Aidelsburger2013}%
  \BibitemOpen
  \bibfield  {author} {\bibinfo {author} {\bibfnamefont {M.}~\bibnamefont
  {Aidelsburger}}, \bibinfo {author} {\bibfnamefont {M.}~\bibnamefont {Atala}},
  \bibinfo {author} {\bibfnamefont {M.}~\bibnamefont {Lohse}}, \bibinfo
  {author} {\bibfnamefont {J.~T.}\ \bibnamefont {Barreiro}}, \bibinfo {author}
  {\bibfnamefont {B.}~\bibnamefont {Paredes}},\ and\ \bibinfo {author}
  {\bibfnamefont {I.}~\bibnamefont {Bloch}},\ }\bibfield  {title} {\bibinfo
  {title} {Realization of the {H}ofstadter {H}amiltonian with ultracold atoms
  in optical lattices},\ }\href
  {https://doi.org/10.1103/PhysRevLett.111.185301} {\bibfield  {journal}
  {\bibinfo  {journal} {Phys. Rev. Lett.}\ }\textbf {\bibinfo {volume} {111}},\
  \bibinfo {pages} {185301} (\bibinfo {year} {2013})}\BibitemShut {NoStop}%
\bibitem [{\citenamefont {Miyake}\ \emph {et~al.}(2013)\citenamefont {Miyake},
  \citenamefont {Siviloglou}, \citenamefont {Kennedy}, \citenamefont {Burton},\
  and\ \citenamefont {Ketterle}}]{Miyake2013}%
  \BibitemOpen
  \bibfield  {author} {\bibinfo {author} {\bibfnamefont {H.}~\bibnamefont
  {Miyake}}, \bibinfo {author} {\bibfnamefont {G.~A.}\ \bibnamefont
  {Siviloglou}}, \bibinfo {author} {\bibfnamefont {C.~J.}\ \bibnamefont
  {Kennedy}}, \bibinfo {author} {\bibfnamefont {W.~C.}\ \bibnamefont
  {Burton}},\ and\ \bibinfo {author} {\bibfnamefont {W.}~\bibnamefont
  {Ketterle}},\ }\bibfield  {title} {\bibinfo {title} {Realizing the {H}arper
  {H}amiltonian with laser-assisted tunneling in optical lattices},\ }\href
  {https://doi.org/10.1103/PhysRevLett.111.185302} {\bibfield  {journal}
  {\bibinfo  {journal} {Phys. Rev. Lett.}\ }\textbf {\bibinfo {volume} {111}},\
  \bibinfo {pages} {185302} (\bibinfo {year} {2013})}\BibitemShut {NoStop}%
\bibitem [{\citenamefont {Struck}\ \emph {et~al.}(2013)\citenamefont {Struck},
  \citenamefont {Weinberg}, \citenamefont {Ölschläger}, \citenamefont
  {Windpassinger}, \citenamefont {Simonet}, \citenamefont {Sengstock},
  \citenamefont {Höppner}, \citenamefont {Hauke}, \citenamefont {Eckardt},
  \citenamefont {Lewenstein},\ and\ \citenamefont {Mathey}}]{Struck2013}%
  \BibitemOpen
  \bibfield  {author} {\bibinfo {author} {\bibfnamefont {J.}~\bibnamefont
  {Struck}}, \bibinfo {author} {\bibfnamefont {M.}~\bibnamefont {Weinberg}},
  \bibinfo {author} {\bibfnamefont {C.}~\bibnamefont {Ölschläger}}, \bibinfo
  {author} {\bibfnamefont {P.}~\bibnamefont {Windpassinger}}, \bibinfo {author}
  {\bibfnamefont {J.}~\bibnamefont {Simonet}}, \bibinfo {author} {\bibfnamefont
  {K.}~\bibnamefont {Sengstock}}, \bibinfo {author} {\bibfnamefont
  {R.}~\bibnamefont {Höppner}}, \bibinfo {author} {\bibfnamefont
  {P.}~\bibnamefont {Hauke}}, \bibinfo {author} {\bibfnamefont
  {A.}~\bibnamefont {Eckardt}}, \bibinfo {author} {\bibfnamefont
  {M.}~\bibnamefont {Lewenstein}},\ and\ \bibinfo {author} {\bibfnamefont
  {L.}~\bibnamefont {Mathey}},\ }\bibfield  {title} {\bibinfo {title}
  {Engineering {I}sing-{XY} spin-models in a triangular lattice using tunable
  artificial gauge fields},\ }\href {https://doi.org/10.1038/nphys2750}
  {\bibfield  {journal} {\bibinfo  {journal} {Nature Physics}\ }\textbf
  {\bibinfo {volume} {9}},\ \bibinfo {pages} {738–743} (\bibinfo {year}
  {2013})}\BibitemShut {NoStop}%
\bibitem [{\citenamefont {Fazzini}\ \emph {et~al.}(2021)\citenamefont
  {Fazzini}, \citenamefont {Chudzinski}, \citenamefont {Dauer}, \citenamefont
  {Schneider},\ and\ \citenamefont {Eggert}}]{prl21}%
  \BibitemOpen
  \bibfield  {author} {\bibinfo {author} {\bibfnamefont {S.}~\bibnamefont
  {Fazzini}}, \bibinfo {author} {\bibfnamefont {P.}~\bibnamefont {Chudzinski}},
  \bibinfo {author} {\bibfnamefont {C.}~\bibnamefont {Dauer}}, \bibinfo
  {author} {\bibfnamefont {I.}~\bibnamefont {Schneider}},\ and\ \bibinfo
  {author} {\bibfnamefont {S.}~\bibnamefont {Eggert}},\ }\bibfield  {title}
  {\bibinfo {title} {Nonequilibrium {F}loquet steady states of time-periodic
  driven {L}uttinger liquids},\ }\href
  {https://doi.org/10.1103/PhysRevLett.126.243401} {\bibfield  {journal}
  {\bibinfo  {journal} {Phys. Rev. Lett.}\ }\textbf {\bibinfo {volume} {126}},\
  \bibinfo {pages} {243401} (\bibinfo {year} {2021})}\BibitemShut {NoStop}%
\bibitem [{\citenamefont {Schweizer}\ \emph {et~al.}(2019)\citenamefont
  {Schweizer}, \citenamefont {Grusdt}, \citenamefont {Berngruber},
  \citenamefont {Barbiero}, \citenamefont {Demler}, \citenamefont {Goldman},
  \citenamefont {Bloch},\ and\ \citenamefont {Aidelsburger}}]{Schweizer2019}%
  \BibitemOpen
  \bibfield  {author} {\bibinfo {author} {\bibfnamefont {C.}~\bibnamefont
  {Schweizer}}, \bibinfo {author} {\bibfnamefont {F.}~\bibnamefont {Grusdt}},
  \bibinfo {author} {\bibfnamefont {M.}~\bibnamefont {Berngruber}}, \bibinfo
  {author} {\bibfnamefont {L.}~\bibnamefont {Barbiero}}, \bibinfo {author}
  {\bibfnamefont {E.}~\bibnamefont {Demler}}, \bibinfo {author} {\bibfnamefont
  {N.}~\bibnamefont {Goldman}}, \bibinfo {author} {\bibfnamefont
  {I.}~\bibnamefont {Bloch}},\ and\ \bibinfo {author} {\bibfnamefont
  {M.}~\bibnamefont {Aidelsburger}},\ }\bibfield  {title} {\bibinfo {title}
  {{{F}loquet approach to Z2 lattice gauge theories with ultracold atoms in
  optical lattices}},\ }\href {https://doi.org/10.1038/s41567-019-0649-7}
  {\bibfield  {journal} {\bibinfo  {journal} {Nature Physics}\ }\textbf
  {\bibinfo {volume} {15}},\ \bibinfo {pages} {1168} (\bibinfo {year}
  {2019})}\BibitemShut {NoStop}%
\bibitem [{\citenamefont {Langen}\ \emph {et~al.}(2015)\citenamefont {Langen},
  \citenamefont {Geiger},\ and\ \citenamefont {Schmiedmayer}}]{Langen2015}%
  \BibitemOpen
  \bibfield  {author} {\bibinfo {author} {\bibfnamefont {T.}~\bibnamefont
  {Langen}}, \bibinfo {author} {\bibfnamefont {R.}~\bibnamefont {Geiger}},\
  and\ \bibinfo {author} {\bibfnamefont {J.}~\bibnamefont {Schmiedmayer}},\
  }\bibfield  {title} {\bibinfo {title} {Ultracold atoms out of equilibrium},\
  }\href {https://doi.org/10.1146/annurev-conmatphys-031214-014548} {\bibfield
  {journal} {\bibinfo  {journal} {Annual Review of Condensed Matter Physics}\
  }\textbf {\bibinfo {volume} {6}},\ \bibinfo {pages} {201} (\bibinfo {year}
  {2015})}\BibitemShut {NoStop}%
\bibitem [{\citenamefont {Gross}\ and\ \citenamefont
  {Bloch}(2017)}]{Gross2017}%
  \BibitemOpen
  \bibfield  {author} {\bibinfo {author} {\bibfnamefont {C.}~\bibnamefont
  {Gross}}\ and\ \bibinfo {author} {\bibfnamefont {I.}~\bibnamefont {Bloch}},\
  }\bibfield  {title} {\bibinfo {title} {Quantum simulations with ultracold
  atoms in optical lattices},\ }\href {https://doi.org/10.1126/science.aal3837}
  {\bibfield  {journal} {\bibinfo  {journal} {Science}\ }\textbf {\bibinfo
  {volume} {357}},\ \bibinfo {pages} {995} (\bibinfo {year}
  {2017})}\BibitemShut {NoStop}%
\bibitem [{\citenamefont {Frölian}\ \emph {et~al.}(2022)\citenamefont
  {Frölian}, \citenamefont {Chisholm}, \citenamefont {Neri}, \citenamefont
  {Cabrera}, \citenamefont {Ramos}, \citenamefont {Celi},\ and\ \citenamefont
  {Tarruell}}]{Froelian2022}%
  \BibitemOpen
  \bibfield  {author} {\bibinfo {author} {\bibfnamefont {A.}~\bibnamefont
  {Frölian}}, \bibinfo {author} {\bibfnamefont {C.~S.}\ \bibnamefont
  {Chisholm}}, \bibinfo {author} {\bibfnamefont {E.}~\bibnamefont {Neri}},
  \bibinfo {author} {\bibfnamefont {C.~R.}\ \bibnamefont {Cabrera}}, \bibinfo
  {author} {\bibfnamefont {R.}~\bibnamefont {Ramos}}, \bibinfo {author}
  {\bibfnamefont {A.}~\bibnamefont {Celi}},\ and\ \bibinfo {author}
  {\bibfnamefont {L.}~\bibnamefont {Tarruell}},\ }\bibfield  {title} {\bibinfo
  {title} {Realizing a 1{D} topological gauge theory in an optically dressed
  bec},\ }\href {https://doi.org/10.1038/s41586-022-04943-3} {\bibfield
  {journal} {\bibinfo  {journal} {Nature}\ }\textbf {\bibinfo {volume} {608}},\
  \bibinfo {pages} {293} (\bibinfo {year} {2022})}\BibitemShut {NoStop}%
\bibitem [{\citenamefont {Chisholm}\ \emph {et~al.}(2022)\citenamefont
  {Chisholm}, \citenamefont {Fr\"olian}, \citenamefont {Neri}, \citenamefont
  {Ramos}, \citenamefont {Tarruell},\ and\ \citenamefont
  {Celi}}]{Chisholm2022}%
  \BibitemOpen
  \bibfield  {author} {\bibinfo {author} {\bibfnamefont {C.~S.}\ \bibnamefont
  {Chisholm}}, \bibinfo {author} {\bibfnamefont {A.}~\bibnamefont {Fr\"olian}},
  \bibinfo {author} {\bibfnamefont {E.}~\bibnamefont {Neri}}, \bibinfo {author}
  {\bibfnamefont {R.}~\bibnamefont {Ramos}}, \bibinfo {author} {\bibfnamefont
  {L.}~\bibnamefont {Tarruell}},\ and\ \bibinfo {author} {\bibfnamefont
  {A.}~\bibnamefont {Celi}},\ }\bibfield  {title} {\bibinfo {title} {Encoding a
  one-dimensional topological gauge theory in a raman-coupled {B}ose-{E}instein
  condensate},\ }\href {https://doi.org/10.1103/PhysRevResearch.4.043088}
  {\bibfield  {journal} {\bibinfo  {journal} {Phys. Rev. Res.}\ }\textbf
  {\bibinfo {volume} {4}},\ \bibinfo {pages} {043088} (\bibinfo {year}
  {2022})}\BibitemShut {NoStop}%
\bibitem [{\citenamefont {Nakamura}(1999)}]{Nakamura99}%
  \BibitemOpen
  \bibfield  {author} {\bibinfo {author} {\bibfnamefont {M.}~\bibnamefont
  {Nakamura}},\ }\bibfield  {title} {\bibinfo {title} {Mechanism of {CDW}-{SDW}
  transition in one dimension},\ }\href {https://doi.org/10.1143/JPSJ.68.3123}
  {\bibfield  {journal} {\bibinfo  {journal} {Journal of the Physical Society
  of Japan}\ }\textbf {\bibinfo {volume} {68}},\ \bibinfo {pages} {3123}
  (\bibinfo {year} {1999})},\ \Eprint
  {https://arxiv.org/abs/https://doi.org/10.1143/JPSJ.68.3123}
  {https://doi.org/10.1143/JPSJ.68.3123} \BibitemShut {NoStop}%
\bibitem [{\citenamefont {Sengupta}\ \emph {et~al.}(2002)\citenamefont
  {Sengupta}, \citenamefont {Sandvik},\ and\ \citenamefont
  {Campbell}}]{Sengupta2002}%
  \BibitemOpen
  \bibfield  {author} {\bibinfo {author} {\bibfnamefont {P.}~\bibnamefont
  {Sengupta}}, \bibinfo {author} {\bibfnamefont {A.~W.}\ \bibnamefont
  {Sandvik}},\ and\ \bibinfo {author} {\bibfnamefont {D.~K.}\ \bibnamefont
  {Campbell}},\ }\bibfield  {title} {\bibinfo {title} {Bond-order-wave phase
  and quantum phase transitions in the one-dimensional extended {H}ubbard
  model},\ }\href {https://doi.org/10.1103/PhysRevB.65.155113} {\bibfield
  {journal} {\bibinfo  {journal} {Phys. Rev. B}\ }\textbf {\bibinfo {volume}
  {65}},\ \bibinfo {pages} {155113} (\bibinfo {year} {2002})}\BibitemShut
  {NoStop}%
\bibitem [{\citenamefont {Sandvik}\ \emph {et~al.}(2004)\citenamefont
  {Sandvik}, \citenamefont {Balents},\ and\ \citenamefont
  {Campbell}}]{Sandvik2004}%
  \BibitemOpen
  \bibfield  {author} {\bibinfo {author} {\bibfnamefont {A.~W.}\ \bibnamefont
  {Sandvik}}, \bibinfo {author} {\bibfnamefont {L.}~\bibnamefont {Balents}},\
  and\ \bibinfo {author} {\bibfnamefont {D.~K.}\ \bibnamefont {Campbell}},\
  }\bibfield  {title} {\bibinfo {title} {Ground state phases of the half-filled
  one-dimensional extended {H}ubbard model},\ }\href
  {https://doi.org/10.1103/PhysRevLett.92.236401} {\bibfield  {journal}
  {\bibinfo  {journal} {Phys. Rev. Lett.}\ }\textbf {\bibinfo {volume} {92}},\
  \bibinfo {pages} {236401} (\bibinfo {year} {2004})}\BibitemShut {NoStop}%
\bibitem [{\citenamefont {Barbiero}\ \emph {et~al.}(2017)\citenamefont
  {Barbiero}, \citenamefont {Fazzini},\ and\ \citenamefont
  {Montorsi}}]{Barbiero2017}%
  \BibitemOpen
  \bibfield  {author} {\bibinfo {author} {\bibfnamefont {L.}~\bibnamefont
  {Barbiero}}, \bibinfo {author} {\bibfnamefont {S.}~\bibnamefont {Fazzini}},\
  and\ \bibinfo {author} {\bibfnamefont {A.}~\bibnamefont {Montorsi}},\
  }\bibfield  {title} {\bibinfo {title} {Non-local order parameters as a probe
  for phase transitions in the extended {F}ermi-{H}ubbard model},\ }\href
  {https://doi.org/https://doi.org/10.1140/epjst/e2016-60386-1} {\bibfield
  {journal} {\bibinfo  {journal} {Eur. Phys. J. Spec. Top.}\ }\textbf {\bibinfo
  {volume} {226}},\ \bibinfo {pages} {2697} (\bibinfo {year}
  {2017})}\BibitemShut {NoStop}%
\bibitem [{\citenamefont {Dalla~Torre}\ \emph {et~al.}(2006)\citenamefont
  {Dalla~Torre}, \citenamefont {Berg},\ and\ \citenamefont
  {Altman}}]{Torre2006}%
  \BibitemOpen
  \bibfield  {author} {\bibinfo {author} {\bibfnamefont {E.~G.}\ \bibnamefont
  {Dalla~Torre}}, \bibinfo {author} {\bibfnamefont {E.}~\bibnamefont {Berg}},\
  and\ \bibinfo {author} {\bibfnamefont {E.}~\bibnamefont {Altman}},\
  }\bibfield  {title} {\bibinfo {title} {Hidden order in 1{D} {B}ose
  insulators},\ }\href {https://doi.org/10.1103/PhysRevLett.97.260401}
  {\bibfield  {journal} {\bibinfo  {journal} {Phys. Rev. Lett.}\ }\textbf
  {\bibinfo {volume} {97}},\ \bibinfo {pages} {260401} (\bibinfo {year}
  {2006})}\BibitemShut {NoStop}%
\bibitem [{\citenamefont {Berg}\ \emph {et~al.}(2008)\citenamefont {Berg},
  \citenamefont {Dalla~Torre}, \citenamefont {Giamarchi},\ and\ \citenamefont
  {Altman}}]{Berg2008}%
  \BibitemOpen
  \bibfield  {author} {\bibinfo {author} {\bibfnamefont {E.}~\bibnamefont
  {Berg}}, \bibinfo {author} {\bibfnamefont {E.~G.}\ \bibnamefont
  {Dalla~Torre}}, \bibinfo {author} {\bibfnamefont {T.}~\bibnamefont
  {Giamarchi}},\ and\ \bibinfo {author} {\bibfnamefont {E.}~\bibnamefont
  {Altman}},\ }\bibfield  {title} {\bibinfo {title} {Rise and fall of hidden
  string order of lattice {B}osons},\ }\href
  {https://doi.org/10.1103/PhysRevB.77.245119} {\bibfield  {journal} {\bibinfo
  {journal} {Phys. Rev. B}\ }\textbf {\bibinfo {volume} {77}},\ \bibinfo
  {pages} {245119} (\bibinfo {year} {2008})}\BibitemShut {NoStop}%
\bibitem [{\citenamefont {Ejima}\ \emph {et~al.}(2014)\citenamefont {Ejima},
  \citenamefont {Lange},\ and\ \citenamefont {Fehske}}]{Ejima2014}%
  \BibitemOpen
  \bibfield  {author} {\bibinfo {author} {\bibfnamefont {S.}~\bibnamefont
  {Ejima}}, \bibinfo {author} {\bibfnamefont {F.}~\bibnamefont {Lange}},\ and\
  \bibinfo {author} {\bibfnamefont {H.}~\bibnamefont {Fehske}},\ }\bibfield
  {title} {\bibinfo {title} {Spectral and entanglement properties of the
  {B}osonic {H}aldane insulator},\ }\href
  {https://doi.org/10.1103/PhysRevLett.113.020401} {\bibfield  {journal}
  {\bibinfo  {journal} {Phys. Rev. Lett.}\ }\textbf {\bibinfo {volume} {113}},\
  \bibinfo {pages} {020401} (\bibinfo {year} {2014})}\BibitemShut {NoStop}%
\bibitem [{\citenamefont {Rossini}\ and\ \citenamefont
  {Fazio}(2012)}]{Rossini2012}%
  \BibitemOpen
  \bibfield  {author} {\bibinfo {author} {\bibfnamefont {D.}~\bibnamefont
  {Rossini}}\ and\ \bibinfo {author} {\bibfnamefont {R.}~\bibnamefont
  {Fazio}},\ }\bibfield  {title} {\bibinfo {title} {Phase diagram of the
  extended {B}ose-{H}ubbard model},\ }\href
  {https://doi.org/10.1088/1367-2630/14/6/065012} {\bibfield  {journal}
  {\bibinfo  {journal} {New Journal of Physics}\ }\textbf {\bibinfo {volume}
  {14}},\ \bibinfo {pages} {065012} (\bibinfo {year} {2012})}\BibitemShut
  {NoStop}%
\bibitem [{\citenamefont {Dalmonte}\ \emph {et~al.}(2011)\citenamefont
  {Dalmonte}, \citenamefont {Di~Dio}, \citenamefont {Barbiero},\ and\
  \citenamefont {Ortolani}}]{Dalmonte2011}%
  \BibitemOpen
  \bibfield  {author} {\bibinfo {author} {\bibfnamefont {M.}~\bibnamefont
  {Dalmonte}}, \bibinfo {author} {\bibfnamefont {M.}~\bibnamefont {Di~Dio}},
  \bibinfo {author} {\bibfnamefont {L.}~\bibnamefont {Barbiero}},\ and\
  \bibinfo {author} {\bibfnamefont {F.}~\bibnamefont {Ortolani}},\ }\bibfield
  {title} {\bibinfo {title} {Homogeneous and inhomogeneous magnetic phases of
  constrained dipolar {B}osons},\ }\href
  {https://doi.org/10.1103/PhysRevB.83.155110} {\bibfield  {journal} {\bibinfo
  {journal} {Phys. Rev. B}\ }\textbf {\bibinfo {volume} {83}},\ \bibinfo
  {pages} {155110} (\bibinfo {year} {2011})}\BibitemShut {NoStop}%
\bibitem [{\citenamefont {Schulz}(1986)}]{Schulz1986}%
  \BibitemOpen
  \bibfield  {author} {\bibinfo {author} {\bibfnamefont {H.~J.}\ \bibnamefont
  {Schulz}},\ }\bibfield  {title} {\bibinfo {title} {Phase diagrams and
  correlation exponents for quantum spin chains of arbitrary spin quantum
  number},\ }\href {https://doi.org/10.1103/PhysRevB.34.6372} {\bibfield
  {journal} {\bibinfo  {journal} {Phys. Rev. B}\ }\textbf {\bibinfo {volume}
  {34}},\ \bibinfo {pages} {6372} (\bibinfo {year} {1986})}\BibitemShut
  {NoStop}%
\bibitem [{Sup()}]{Supp}%
  \BibitemOpen
  \href@noop {} {}\bibinfo {note} {In the Appendix details about
  the lattice gauge function, about {B}osonization, about symmetry
  transformations, and additional numerical data are presented.}\BibitemShut
  {Stop}%
\bibitem [{\citenamefont {Lecheminant}\ \emph {et~al.}(2002)\citenamefont
  {Lecheminant}, \citenamefont {Gogolin},\ and\ \citenamefont
  {Nersesyan}}]{Lecheminant2002}%
  \BibitemOpen
  \bibfield  {author} {\bibinfo {author} {\bibfnamefont {P.}~\bibnamefont
  {Lecheminant}}, \bibinfo {author} {\bibfnamefont {A.~O.}\ \bibnamefont
  {Gogolin}},\ and\ \bibinfo {author} {\bibfnamefont {A.~A.}\ \bibnamefont
  {Nersesyan}},\ }\bibfield  {title} {\bibinfo {title} {Criticality in
  self-dual sine-{G}ordon models},\ }\href
  {https://doi.org/https://doi.org/10.1016/S0550-3213(02)00474-1} {\bibfield
  {journal} {\bibinfo  {journal} {Nuclear Physics B}\ }\textbf {\bibinfo
  {volume} {639}},\ \bibinfo {pages} {502} (\bibinfo {year}
  {2002})}\BibitemShut {NoStop}%
\bibitem [{\citenamefont {White}(1992)}]{White1992}%
  \BibitemOpen
  \bibfield  {author} {\bibinfo {author} {\bibfnamefont {S.~R.}\ \bibnamefont
  {White}},\ }\bibfield  {title} {\bibinfo {title} {Density matrix formulation
  for quantum renormalization groups},\ }\href
  {https://doi.org/10.1103/PhysRevLett.69.2863} {\bibfield  {journal} {\bibinfo
   {journal} {Phys. Rev. Lett.}\ }\textbf {\bibinfo {volume} {69}},\ \bibinfo
  {pages} {2863} (\bibinfo {year} {1992})}\BibitemShut {NoStop}%
\bibitem [{\citenamefont {White}(1993)}]{White1993}%
  \BibitemOpen
  \bibfield  {author} {\bibinfo {author} {\bibfnamefont {S.~R.}\ \bibnamefont
  {White}},\ }\bibfield  {title} {\bibinfo {title} {Density-matrix algorithms
  for quantum renormalization groups},\ }\href
  {https://doi.org/10.1103/PhysRevB.48.10345} {\bibfield  {journal} {\bibinfo
  {journal} {Phys. Rev. B}\ }\textbf {\bibinfo {volume} {48}},\ \bibinfo
  {pages} {10345} (\bibinfo {year} {1993})}\BibitemShut {NoStop}%
\bibitem [{\citenamefont {McCulloch}(2008)}]{Mcculloch2008}%
  \BibitemOpen
  \bibfield  {author} {\bibinfo {author} {\bibfnamefont {I.~P.}\ \bibnamefont
  {McCulloch}},\ }\href@noop {} {\bibinfo {title} {Infinite size density matrix
  renormalization group, revisited}} (\bibinfo {year} {2008}),\ \Eprint
  {https://arxiv.org/abs/0804.2509} {arXiv:0804.2509 [cond-mat.str-el]}
  \BibitemShut {NoStop}%
\bibitem [{\citenamefont {Ogino}\ \emph {et~al.}(2021)\citenamefont {Ogino},
  \citenamefont {Furukawa}, \citenamefont {Kaneko}, \citenamefont {Morita},\
  and\ \citenamefont {Kawashima}}]{Ogino2021}%
  \BibitemOpen
  \bibfield  {author} {\bibinfo {author} {\bibfnamefont {T.}~\bibnamefont
  {Ogino}}, \bibinfo {author} {\bibfnamefont {S.}~\bibnamefont {Furukawa}},
  \bibinfo {author} {\bibfnamefont {R.}~\bibnamefont {Kaneko}}, \bibinfo
  {author} {\bibfnamefont {S.}~\bibnamefont {Morita}},\ and\ \bibinfo {author}
  {\bibfnamefont {N.}~\bibnamefont {Kawashima}},\ }\bibfield  {title} {\bibinfo
  {title} {Symmetry-protected topological phases and competing orders in a
  spin-$\frac{1}{2}$ xxz ladder with a four-spin interaction},\ }\href
  {https://doi.org/10.1103/PhysRevB.104.075135} {\bibfield  {journal} {\bibinfo
   {journal} {Phys. Rev. B}\ }\textbf {\bibinfo {volume} {104}},\ \bibinfo
  {pages} {075135} (\bibinfo {year} {2021})}\BibitemShut {NoStop}%
\bibitem [{\citenamefont {Nishimoto}(2011)}]{Nishimoto2011}%
  \BibitemOpen
  \bibfield  {author} {\bibinfo {author} {\bibfnamefont {S.}~\bibnamefont
  {Nishimoto}},\ }\bibfield  {title} {\bibinfo {title}
  {Tomonaga-{L}uttinger-liquid criticality: Numerical entanglement entropy
  approach},\ }\href {https://doi.org/10.1103/PhysRevB.84.195108} {\bibfield
  {journal} {\bibinfo  {journal} {Phys. Rev. B}\ }\textbf {\bibinfo {volume}
  {84}},\ \bibinfo {pages} {195108} (\bibinfo {year} {2011})}\BibitemShut
  {NoStop}%
\bibitem [{\citenamefont {Mac\^edo}\ \emph {et~al.}(2022)\citenamefont
  {Mac\^edo}, \citenamefont {Ramos},\ and\ \citenamefont
  {Pereira}}]{Macedo2022}%
  \BibitemOpen
  \bibfield  {author} {\bibinfo {author} {\bibfnamefont {R.~A.}\ \bibnamefont
  {Mac\^edo}}, \bibinfo {author} {\bibfnamefont {F.~B.}\ \bibnamefont
  {Ramos}},\ and\ \bibinfo {author} {\bibfnamefont {R.~G.}\ \bibnamefont
  {Pereira}},\ }\bibfield  {title} {\bibinfo {title} {Continuous phase
  transition from a chiral spin state to collinear magnetic order in a zigzag
  chain with {K}itaev interactions},\ }\href
  {https://doi.org/10.1103/PhysRevB.105.205144} {\bibfield  {journal} {\bibinfo
   {journal} {Phys. Rev. B}\ }\textbf {\bibinfo {volume} {105}},\ \bibinfo
  {pages} {205144} (\bibinfo {year} {2022})}\BibitemShut {NoStop}%
\bibitem [{\citenamefont {Mudry}\ \emph {et~al.}(2019)\citenamefont {Mudry},
  \citenamefont {Furusaki}, \citenamefont {Morimoto},\ and\ \citenamefont
  {Hikihara}}]{Mudry2019}%
  \BibitemOpen
  \bibfield  {author} {\bibinfo {author} {\bibfnamefont {C.}~\bibnamefont
  {Mudry}}, \bibinfo {author} {\bibfnamefont {A.}~\bibnamefont {Furusaki}},
  \bibinfo {author} {\bibfnamefont {T.}~\bibnamefont {Morimoto}},\ and\
  \bibinfo {author} {\bibfnamefont {T.}~\bibnamefont {Hikihara}},\ }\bibfield
  {title} {\bibinfo {title} {Quantum phase transitions beyond
  {L}andau-{G}inzburg theory in one-dimensional space revisited},\ }\href
  {https://doi.org/10.1103/PhysRevB.99.205153} {\bibfield  {journal} {\bibinfo
  {journal} {Phys. Rev. B}\ }\textbf {\bibinfo {volume} {99}},\ \bibinfo
  {pages} {205153} (\bibinfo {year} {2019})}\BibitemShut {NoStop}%
\bibitem [{\citenamefont {Jiang}\ and\ \citenamefont
  {Motrunich}(2019)}]{Jiang2019}%
  \BibitemOpen
  \bibfield  {author} {\bibinfo {author} {\bibfnamefont {S.}~\bibnamefont
  {Jiang}}\ and\ \bibinfo {author} {\bibfnamefont {O.}~\bibnamefont
  {Motrunich}},\ }\bibfield  {title} {\bibinfo {title} {{I}sing ferromagnet to
  valence bond solid transition in a one-dimensional spin chain: Analogies to
  deconfined quantum critical points},\ }\href
  {https://doi.org/10.1103/PhysRevB.99.075103} {\bibfield  {journal} {\bibinfo
  {journal} {Phys. Rev. B}\ }\textbf {\bibinfo {volume} {99}},\ \bibinfo
  {pages} {075103} (\bibinfo {year} {2019})}\BibitemShut {NoStop}%
\bibitem [{\citenamefont {Huang}\ \emph {et~al.}(2019)\citenamefont {Huang},
  \citenamefont {Lu}, \citenamefont {You}, \citenamefont {Meng},\ and\
  \citenamefont {Xiang}}]{Huang2019}%
  \BibitemOpen
  \bibfield  {author} {\bibinfo {author} {\bibfnamefont {R.-Z.}\ \bibnamefont
  {Huang}}, \bibinfo {author} {\bibfnamefont {D.-C.}\ \bibnamefont {Lu}},
  \bibinfo {author} {\bibfnamefont {Y.-Z.}\ \bibnamefont {You}}, \bibinfo
  {author} {\bibfnamefont {Z.~Y.}\ \bibnamefont {Meng}},\ and\ \bibinfo
  {author} {\bibfnamefont {T.}~\bibnamefont {Xiang}},\ }\bibfield  {title}
  {\bibinfo {title} {Emergent symmetry and conserved current at a
  one-dimensional incarnation of deconfined quantum critical point},\ }\href
  {https://doi.org/10.1103/PhysRevB.100.125137} {\bibfield  {journal} {\bibinfo
   {journal} {Phys. Rev. B}\ }\textbf {\bibinfo {volume} {100}},\ \bibinfo
  {pages} {125137} (\bibinfo {year} {2019})}\BibitemShut {NoStop}%
\bibitem [{\citenamefont {Roberts}\ \emph {et~al.}(2019)\citenamefont
  {Roberts}, \citenamefont {Jiang},\ and\ \citenamefont
  {Motrunich}}]{Roberts2019}%
  \BibitemOpen
  \bibfield  {author} {\bibinfo {author} {\bibfnamefont {B.}~\bibnamefont
  {Roberts}}, \bibinfo {author} {\bibfnamefont {S.}~\bibnamefont {Jiang}},\
  and\ \bibinfo {author} {\bibfnamefont {O.~I.}\ \bibnamefont {Motrunich}},\
  }\bibfield  {title} {\bibinfo {title} {Deconfined quantum critical point in
  one dimension},\ }\href {https://doi.org/10.1103/PhysRevB.99.165143}
  {\bibfield  {journal} {\bibinfo  {journal} {Phys. Rev. B}\ }\textbf {\bibinfo
  {volume} {99}},\ \bibinfo {pages} {165143} (\bibinfo {year}
  {2019})}\BibitemShut {NoStop}%
\bibitem [{\citenamefont {Luo}\ \emph {et~al.}(2023)\citenamefont {Luo},
  \citenamefont {Hu}, \citenamefont {Li}, \citenamefont {Zhao}, \citenamefont
  {Kee},\ and\ \citenamefont {Wang}}]{Luo2023}%
  \BibitemOpen
  \bibfield  {author} {\bibinfo {author} {\bibfnamefont {Q.}~\bibnamefont
  {Luo}}, \bibinfo {author} {\bibfnamefont {S.}~\bibnamefont {Hu}}, \bibinfo
  {author} {\bibfnamefont {J.}~\bibnamefont {Li}}, \bibinfo {author}
  {\bibfnamefont {J.}~\bibnamefont {Zhao}}, \bibinfo {author} {\bibfnamefont
  {H.-Y.}\ \bibnamefont {Kee}},\ and\ \bibinfo {author} {\bibfnamefont
  {X.}~\bibnamefont {Wang}},\ }\bibfield  {title} {\bibinfo {title}
  {Spontaneous dimerization, spin-nematic order, and deconfined quantum
  critical point in a spin-1 {K}itaev chain with tunable single-ion
  anisotropy},\ }\href {https://doi.org/10.1103/PhysRevB.107.245131} {\bibfield
   {journal} {\bibinfo  {journal} {Phys. Rev. B}\ }\textbf {\bibinfo {volume}
  {107}},\ \bibinfo {pages} {245131} (\bibinfo {year} {2023})}\BibitemShut
  {NoStop}%
\bibitem [{\citenamefont {Zhang}\ and\ \citenamefont
  {Levin}(2023)}]{Zhang2023}%
  \BibitemOpen
  \bibfield  {author} {\bibinfo {author} {\bibfnamefont {C.}~\bibnamefont
  {Zhang}}\ and\ \bibinfo {author} {\bibfnamefont {M.}~\bibnamefont {Levin}},\
  }\bibfield  {title} {\bibinfo {title} {Exactly solvable model for a
  deconfined quantum critical point in 1{D}},\ }\href
  {https://doi.org/10.1103/PhysRevLett.130.026801} {\bibfield  {journal}
  {\bibinfo  {journal} {Phys. Rev. Lett.}\ }\textbf {\bibinfo {volume} {130}},\
  \bibinfo {pages} {026801} (\bibinfo {year} {2023})}\BibitemShut {NoStop}%
\bibitem [{\citenamefont {Senthil}\ \emph {et~al.}(2004)\citenamefont
  {Senthil}, \citenamefont {Balents}, \citenamefont {Sachdev}, \citenamefont
  {Vishwanath},\ and\ \citenamefont {Fisher}}]{Senthil2004}%
  \BibitemOpen
  \bibfield  {author} {\bibinfo {author} {\bibfnamefont {T.}~\bibnamefont
  {Senthil}}, \bibinfo {author} {\bibfnamefont {L.}~\bibnamefont {Balents}},
  \bibinfo {author} {\bibfnamefont {S.}~\bibnamefont {Sachdev}}, \bibinfo
  {author} {\bibfnamefont {A.}~\bibnamefont {Vishwanath}},\ and\ \bibinfo
  {author} {\bibfnamefont {M.~P.~A.}\ \bibnamefont {Fisher}},\ }\bibfield
  {title} {\bibinfo {title} {Quantum criticality beyond the
  {L}andau-{G}inzburg-wilson paradigm},\ }\href
  {https://doi.org/10.1103/PhysRevB.70.144407} {\bibfield  {journal} {\bibinfo
  {journal} {Phys. Rev. B}\ }\textbf {\bibinfo {volume} {70}},\ \bibinfo
  {pages} {144407} (\bibinfo {year} {2004})}\BibitemShut {NoStop}%
\bibitem [{\citenamefont {Levin}\ and\ \citenamefont
  {Senthil}(2004)}]{Levin2004}%
  \BibitemOpen
  \bibfield  {author} {\bibinfo {author} {\bibfnamefont {M.}~\bibnamefont
  {Levin}}\ and\ \bibinfo {author} {\bibfnamefont {T.}~\bibnamefont
  {Senthil}},\ }\bibfield  {title} {\bibinfo {title} {Deconfined quantum
  criticality and n\'eel order via dimer disorder},\ }\href
  {https://doi.org/10.1103/PhysRevB.70.220403} {\bibfield  {journal} {\bibinfo
  {journal} {Phys. Rev. B}\ }\textbf {\bibinfo {volume} {70}},\ \bibinfo
  {pages} {220403} (\bibinfo {year} {2004})}\BibitemShut {NoStop}%
\bibitem [{\citenamefont {Wang}\ \emph {et~al.}(2017)\citenamefont {Wang},
  \citenamefont {Nahum}, \citenamefont {Metlitski}, \citenamefont {Xu},\ and\
  \citenamefont {Senthil}}]{Wang2017}%
  \BibitemOpen
  \bibfield  {author} {\bibinfo {author} {\bibfnamefont {C.}~\bibnamefont
  {Wang}}, \bibinfo {author} {\bibfnamefont {A.}~\bibnamefont {Nahum}},
  \bibinfo {author} {\bibfnamefont {M.~A.}\ \bibnamefont {Metlitski}}, \bibinfo
  {author} {\bibfnamefont {C.}~\bibnamefont {Xu}},\ and\ \bibinfo {author}
  {\bibfnamefont {T.}~\bibnamefont {Senthil}},\ }\bibfield  {title} {\bibinfo
  {title} {Deconfined quantum critical points: Symmetries and dualities},\
  }\href {https://doi.org/10.1103/PhysRevX.7.031051} {\bibfield  {journal}
  {\bibinfo  {journal} {Phys. Rev. X}\ }\textbf {\bibinfo {volume} {7}},\
  \bibinfo {pages} {031051} (\bibinfo {year} {2017})}\BibitemShut {NoStop}%
\bibitem [{\citenamefont {{L}andau}\ and\ \citenamefont
  {Lifshitz}(1980)}]{LandauLifshitz}%
  \BibitemOpen
  \bibfield  {author} {\bibinfo {author} {\bibfnamefont {L.}~\bibnamefont
  {{L}andau}}\ and\ \bibinfo {author} {\bibfnamefont {E.}~\bibnamefont
  {Lifshitz}},\ }\href@noop {} {\emph {\bibinfo {title} {Statistical Physics,
  Part I, Vol.5}}}\ (\bibinfo  {publisher} {Elsevier Butterworth-Heinemann,
  Oxford},\ \bibinfo {year} {1980})\BibitemShut {NoStop}%
\bibitem [{\citenamefont {Ejima}\ and\ \citenamefont
  {Fehske}(2015)}]{Ejima2015}%
  \BibitemOpen
  \bibfield  {author} {\bibinfo {author} {\bibfnamefont {S.}~\bibnamefont
  {Ejima}}\ and\ \bibinfo {author} {\bibfnamefont {H.}~\bibnamefont {Fehske}},\
  }\bibfield  {title} {\bibinfo {title} {Comparative density-matrix
  renormalization group study of symmetry-protected topological phases in
  spin-1 chain and {B}ose-{H}ubbard models},\ }\href
  {https://doi.org/10.1103/PhysRevB.91.045121} {\bibfield  {journal} {\bibinfo
  {journal} {Phys. Rev. B}\ }\textbf {\bibinfo {volume} {91}},\ \bibinfo
  {pages} {045121} (\bibinfo {year} {2015})}\BibitemShut {NoStop}%
\bibitem [{\citenamefont {Giamarchi}(2003)}]{Giamarchi2003}%
  \BibitemOpen
  \bibfield  {author} {\bibinfo {author} {\bibfnamefont {T.}~\bibnamefont
  {Giamarchi}},\ }\href@noop {} {\emph {\bibinfo {title} {Quantum Physics in
  One Dimension}}}\ (\bibinfo  {publisher} {Clarendon Press},\ \bibinfo {year}
  {2003})\BibitemShut {NoStop}%
\end{thebibliography}%

\end{document}